\documentclass[prd,showpacs,twocolumn,superscriptaddress,floatfix,nofootinbib,10pt]{revtex4-2}
\usepackage{setspace}
\usepackage[utf8]{inputenc}
\usepackage{float}
\usepackage{amsmath,amssymb,amsfonts,bm}
\usepackage{graphicx}
\usepackage{multirow}
\usepackage[usenames,dvipsnames]{color}
\allowdisplaybreaks
\usepackage[
colorlinks=true,
linkcolor=blue,
breaklinks=true,
urlcolor=blue,
citecolor=blue]{hyperref}

\usepackage{orcidlink}
\usepackage{subfigure}
\usepackage{soul}
\usepackage{xcolor}
\usepackage{physics}
\usepackage{booktabs}
\usepackage{ulem}
\usepackage{tikz-feynman}
\usepackage{makecell}
\definecolor{green}{rgb}{0,0.6,0}
\newcommand{\mev}{\textrm{ MeV}}

\newcommand{\GXNU}{\affiliation{Department of Physics, Guangxi Normal University, Guilin 541004, China}}

\newcommand{\GXZD}{\affiliation{Guangxi Key Laboratory of Nuclear Physics and Technology, Guangxi Normal University, Guilin 541004, China}}

\newcommand{\NNNU}{\affiliation{School of Physics and Electronics, Nanning Normal University, Nanning 530100, China}}

\newcommand{\HNU}{\affiliation{Center for Theoretical Physics, School of Physics and Optoelectronic Engineering, Hainan University, Haikou 570228, China}}

\newcommand{\IFIC}{\affiliation{Departamento de F\'{\i}sica Te\'orica and IFIC, Centro Mixto Universidad de
		Valencia-CSIC Institutos de Investigaci\'on de Paterna, Apartado 22085,
		46071 Valencia, Spain}}

\begin{document}
	\title{Interaction and correlation functions for $\pi f_1(1285)$, $\eta f_1(1285)$}
	
	\begin{abstract}
	We have studied the interaction of $\pi^0 (\eta) f_1(1285)$ assuming the $f_1(1285)$ to be a molecular state of $K^* \bar K - \bar K^* K$. We use a framework in which a $\pi^0 (\eta) f_1(1285)$ optical potential is obtained, which is later used as the kernel of the Lippmann-Schwinger equation, following the standard method for the interaction of particles with nuclei. 
	The optical potential is obtained using the fixed center approximation to the Faddeev equations, where a cluster, here the $f_1(1285)$, remains unchanged during the interaction, appropriate to the situation that one has here. We have obtained the scattering matrix for this system, the scattering length and effective range, plus the correlation functions. The framework used has been previously tested in the study of the $p  f_1(1285)$ interaction and has been shown to give results in agreement with the recent experimental measurement of the  $p  f_1(1285)$ correlation function. On the other hand, from this interaction we do not obtain clear signals for the $\pi_1(1400)$ or $\pi_1(1600)$, nor for the $\eta_1(1855)$ resonances, which in other approaches have been claimed to arise from the same dynamics. We, however, obtain a structure in the $\pi^0 f_1(1285)$ amplitude around $1500-1600\mev$ and a strong cusp at the $\eta f_1(1285)$ threshold of $1833\mev$.
	\end{abstract}
	
	\author{Wen-Hao Jia\orcidlink{0009-0001-1170-3540}}
	\GXNU%
	\GXZD%

	\author{Hai-Peng Li\orcidlink{0009-0008-2985-3011}}
	\GXNU%
	\NNNU%

	\author{Wei-Hong Liang\orcidlink{0000-0001-5847-2498}}%
	\email{liangwh@gxnu.edu.cn}
	\GXNU%
	\GXZD%

	\author{Jing Song\orcidlink{0000-0003-3789-7504}}
	\HNU%
	\GXNU%

	\author{Eulogio Oset\orcidlink{0000-0002-4462-7919}}%
	\email{Oset@ific.uv.es}
	\GXNU%
	\IFIC%

\maketitle

\section{Introduction}\label{sec:Intr}
The BESIII Collaboration reported the observation of a state branded as $\eta_1(1855)$, with quantum numbers $I^G (J^{PC})=0^+(1^{-+})$ \cite{BESIII:2022iwi}, which make this state exotic from the $q \bar q$ point of view.  
The state was observed in the $J/\psi \to \gamma \eta_1(1855);\;\eta_1(1855) \to \eta \eta'$ decay, and is already tabulated in the Particle Data Group (PDG) \cite{ParticleDataGroup:2024cfk} with only the BESIII entry. 
This state would be related to the $\pi_1(1600)$ with $1^-(1^{-+})$, which, according to the PDG, could also be the $\pi_1(1400)$ that is also tabulated there. 
Theoretical interest in these states is great, since they challenge the conventional $q \bar q$ structure of the mesons and require at least four quarks. 
Even then, there can be different structures like compact tetraquarks or meson-meson molecules. 
In Ref.~\cite{Zhang:2001sb}, the possibility that the $\pi_1(1400)$ and $\pi_1(1600)$ could correspond to molecular states of  $\pi \eta(1295)$ and $\pi \eta(1440)$, respectively, was studied, with negative conclusions. 
In Ref.~\cite{Bernard:2003jd}, a tetraquark nature is advocated with different SU(3) classifications. 
A hybrid nature  for the $\pi_1(1600)$ is advocated in Refs.~\cite{Close:2003af,Eshraim:2020ucw,Li:2021fwk}, but different conclusions are obtained in the work of Ref.~\cite{General:2007bk}. 
Lattice QCD calculations have been able to reproduce a $(1^{-+})$ state compatible with the $\pi_1(1600)$ \cite{Hedditch:2005zf}. 
QCD sum rules have also been applied to obtain a tetraquark compatible with the $\pi_1(1600)$ in Refs.~\cite{Chen:2008qw,Narison:2009vj,Zhang:2013rya}. 
A very different proposal is made in Ref.~\cite{Zhang:2016bmy}, where the $\pi_1(1600)$ is assumed to be a three-body molecular state of $\pi K^* \bar K$. 
Review papers on this issue can be found in Refs.~\cite{Meyer:2015eta,Chen:2022asf}.

The discovery of the $\eta_1(1855)$ state has also raised interest in the theoretical community. 
In Refs.~\cite{Dong:2022cuw,Yang:2022rck,Huang:2022tpq,Liu:2024lph}, it is obtained as a $K \bar K_1(1400)$ molecule. 
Once again the hybrid picture gets support in Refs.~\cite{Shastry:2022mhk,Chen:2023ukh,Tan:2024grd,Qiu:2022ktc}, but in Ref.~\cite{Zhang:2025xee} this structure is shown to lead to widths that are too large compared with experiment. 
A tetraquark structure using QCD sum rules is advocated in Ref.~\cite{Wan:2022xkx}.
  
A different structure for the $\pi_1(1600)$ and $\eta_1(1855)$ states is advocated in Ref.~\cite{Yan:2023vbh}, where the interaction of $\pi$ with $f_1(1285)$, together with other coupled channels of a pseudoscalar meson and an axial vector meson, gives rise to the $\pi_1(1600)$, while the $\eta f_1(1285)$ together with other coupled channels gives rise to the $\eta_1(1855)$. 
The kernel of the interaction is obtained by assuming the axial-vector mesons as matter fields, which provides the Weinberg-Tomozawa interaction of order $O(1/f^2)$, with $f$ being the pion decay constant. 
This work shares some overlap with the one we present here. 
The difference is that we do not consider the axial vector mesons as matter fields but rather as molecular states composed of a pseudoscalar meson and a vector meson \cite{Lutz:2003fm,Roca:2005nm,Zhou:2014ila}. 
We then consider the interaction of a pseudoscalar meson with a molecular state, analogous to the interaction of a particle with a nucleus composed of two particles, where the interaction goes beyond the $O(1/f^2)$ approximation, which anticipates that we should obtain different results. 
  
The interaction of a particle with a nucleus is most commonly addressed by defining an optical potential, and then solving the Schroedinger equation (or the Lippmann-Schwinger equation) \cite{Ericson:1988gk,Seki:1983sh,Nieves:1993ev,Brown:1975di}. 
For the case of a particle interacting with a bound state of two particles, this would involve the construction of an optical potential and the posterior solution of the Lippmann-Schwinger equation. 
This can be done by using the Fixed Center Approximation (FCA) to the Faddeev equations for the three-body system, in which the molecular state is considered as a cluster and the external particle collides with it without breaking it. 
This is the basic assumption in the FCA, which is suited to the present case, where the cluster, the molecule, is identified in both the initial and final states. 
  
The study of particle interactions with axial vector mesons has gained renewed interest following the measurement of the proton-$f_1(1285)$ correlation function reported in Ref.~\cite{Lauraser}. 
In anticipation of this experiment, a calculation was performed in Ref.~\cite{Encarnacion:2025lyf} to obtain the correlation function for this system using the ordinary FCA, in which the last step mentioned above of the solution of the Lippmann-Schwinger equation was not done. Even then, the predicted results for the correlation function are in qualitative agreement with the recently measured magnitude. 
  
The work of Ref.~\cite{Encarnacion:2025lyf} spotted a problem in the FCA method, showing that it did not fulfill elastic unitarity at the particle-cluster threshold. 
Although an empirical solution was proposed in Ref.~\cite{Encarnacion:2025lyf}, a formal resolution was later provided by
Ref.~\cite{Ikeno:2025bsx} in the study of the $n \bar D_{s0}(2317)$ interaction, where the FCA results were understood as a particle-cluster optical potential, analogous to that in particle-nucleus interaction, which was unitarized in the particle-cluster system by means of the solution of the Lippmann-Schwinger equation. 
Simplified formulas for the general solution provided in Ref.~\cite{Ikeno:2025bsx} were found in Ref.~\cite{Agatao:2025ckp}, in the study of the correlation functions for the $n\,\bar{D}_{s1}(2460)$ and $n\,\bar{D}_{s1}(2536)$ systems, where an analytical proof of the elastic unitarity was made.
Further work was done in the study of the super exotic three-body system $K^{*+} D^{*+} K^{*+}$, implementing some off-shell dependence in the formalism through $\Theta\left(q_{\mathrm{max}}-|\vec{q}\,|\right)$ factors that appear in the elementary amplitudes, which are also considered here. 
This is the final status of the evolution of the method, which has been also applied to evaluate correlation functions of the $K f_1(1285)$ in Ref.~\cite{Jia:2026dpl} and $K D_{s0}(2317)$ in Ref.~\cite{Jia:2026iqo}. 
  
The reliability of the method to study the interaction of these systems has been corroborated by the experimental results in Ref.~\cite{Lauraser} for the correlation function of $p f_1(1285)$, which show good agreement with the updated results of Ref.~\cite{Encarnacion:2025lyf} in Ref.~\cite{Encarnacion:2026zas}, in light of the recent developments implementing elastic unitarity, which were obtained prior to the experimental results of Ref.~\cite{Lauraser}.

The paper proceeds as follows: in section \ref{sec:form} we describe the formalism used to study the $\pi^0$ and $\eta$ interaction with the $f_1(1285)$ resonance; in section \ref{sec:res} we show the results and in section \ref{sec:con} we make a discussion of the results and present our conclusions; in section \ref{sec:uncertainties} we discuss the uncertainties in our calculations.

\section{Formalism}\label{sec:form}
We follow closely the work of Ref.~\cite{Jia:2026dpl} for the $K f_1(1285)$ interaction.

In the molecular picture, the $f_1(1285)$, with the phase conventions for the isospin multiplets $(K^+,\,K^0)$, $(\bar{K}^0,\,-K^-)$, $(K^{*+},\,K^{*0})$ and $(\bar{K}^{*0},\,-K^{*-})$, is given by \cite{Roca:2005nm}
\begin{align}
	f_1(1285)&\equiv -\sqrt{\frac{1}{2}}\left(
	K^{*+}K^- + K^{*0}\bar{K}^0 - K^{*-}K^+ - \bar{K}^{*0}K^0
	\right)\notag\\[1.5mm]
	&= \sqrt{\frac{1}{2}}\left[
	\left(K^*\bar{K}\right)^{I=0} + \left(\bar K^* K\right)^{I=0}
	\right].
	\label{eq:1}
\end{align}
One has to differentiate between the building blocks and decay channels. 
The latter ones contribute to the width of the state because they have phase space available. 
However, while they can be incorporated as components in the wave function, their strength is too small to have any significant role in the building up and the mass of the states, and they can be treated perturbatively. 
This is discussed in different works, like those in Refs.~\cite{Molina:2008jw,Geng:2008gx}.
In particular, from this molecular  perspective, the different decay widths of the $f_1(1285)$ have been evaluated: 
the decays into $a_0(980) \pi$, $f_0(980) \pi$ and $K\bar K \pi$ have been successfully tested in Refs.~\cite{Aceti:2015zva,Aceti:2015pma}, and further supporting evidence has come from studies of various reactions, including $K^- p \to f_1(1285) \Lambda$ \cite{Xie:2015wja}, $J/\psi \to \phi f_1(1285)$ \cite{Xie:2015lta}, $B^0_s \to J/\psi f_1(1285)$ \cite{Molina:2016pbg}, $\tau \to f_1(1285) \pi \nu_\tau$ \cite{Oset:2018zgc}, and $\bar B^0 \to J/\psi f_1(1285)$ \cite{He:2021exv}.

We will be studying the interaction of $\pi^0$ and $\eta$ with the $f_1(1285)$. Since $\pi^0$ and $\eta$ are identical to their antiparticles, the interaction of $\pi^0$ or $\eta$ with either of the two components of Eq.~\eqref{eq:1} is the same.
This allows us to concentrate on just one component, namely $\pi^0$ or $\eta$ interacting with $(K^* \bar{K})^{I=0}$.

\begin{figure}[b]
	\begin{center}
		\includegraphics[width=0.48\textwidth]{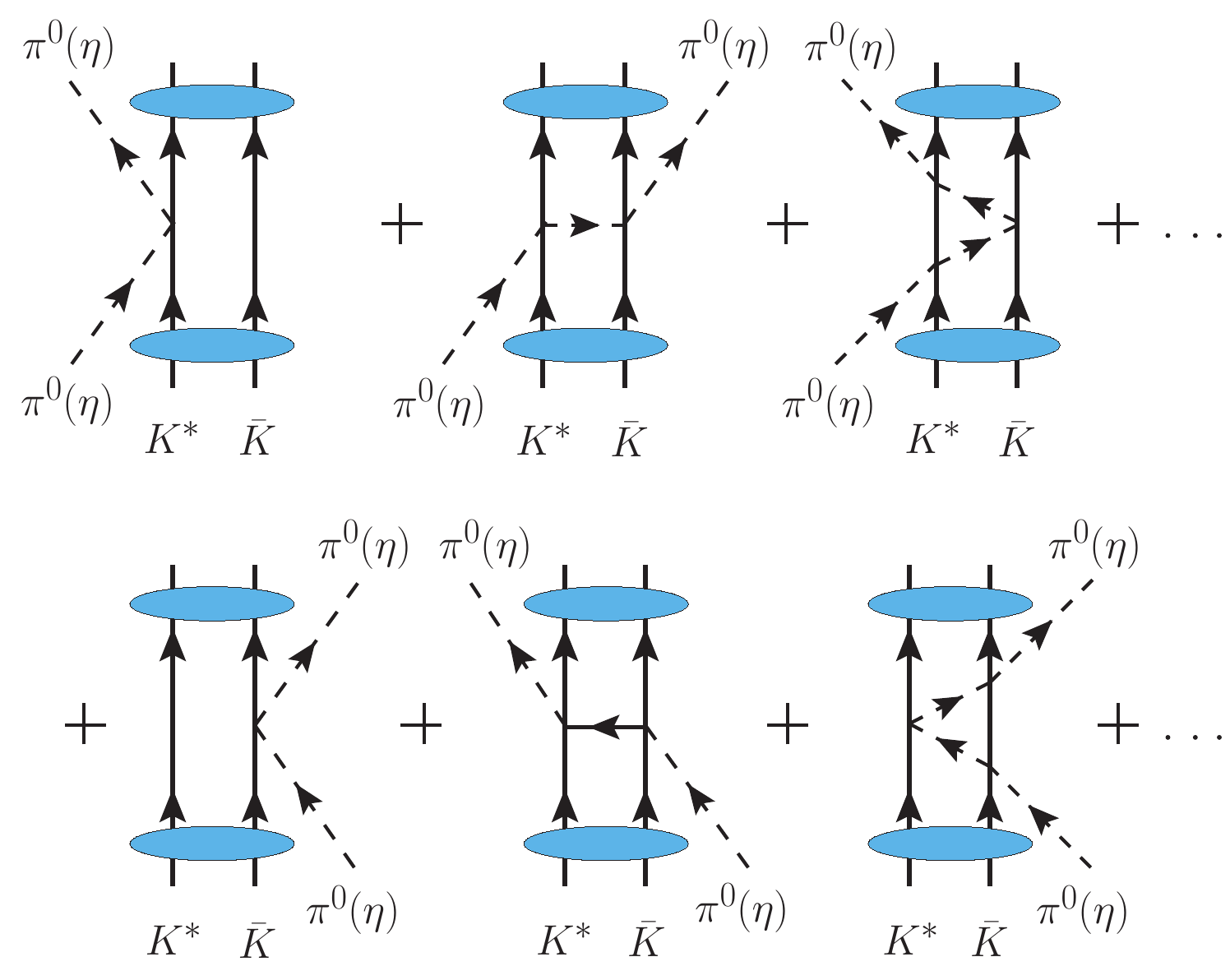}
	\end{center}
	\vspace{-0.5cm}
	\caption{Diagrams of the ordinary FCA for the interaction of $\pi^0$ or $\eta$ with the $K^* \bar{K}$ component of the $f_1(1285)$.}
	\label{Fig1}
\end{figure}

The FCA accounts for the diagrams of Fig.~\ref{Fig1}.
For reasons of normalization of the scattering matrix for the $\pi^0(\eta)$ $f_1(1285)$ interaction, rather than $\pi^0(\eta)$ scattering with $K^*$ or $\bar{K}$, we write the amplitudes as
\begin{equation}
	\tilde{t}_1 = \frac{M_c}{m_{K^*}}t_1,
	\qquad
	\tilde{t}_2 = \frac{M_c}{m_{\bar{K}	}}t_2,
	\label{eq:2}
\end{equation}
where $M_c$ is the mass of the cluster, i.e., the $f_1(1285)$ state.
The matrices $t_1$ and $t_2$ correspond to the scattering of the $\pi^0$ and $\eta$ with the $K^*$ and $\bar{K}$, respectively. One has to consider that $K^* \bar{K}$ is coupled to $I=0$, in which case (see details in Ref.~\cite{Roca:2010tf}), one easily obtains:

$a)$ $\pi^0$ scattering
\begin{equation}
\begin{aligned}
	t_1 &= \dfrac{2}{3}\;t_{\pi K^{*}}^{I=3/2}+\dfrac{1}{3}\;t_{\pi K^{*}}^{I=1/2},\\[2mm]
	t_2 &= \dfrac{2}{3}\;t_{\pi K}^{I=3/2}+\dfrac{1}{3}\;t_{\pi K}^{I=1/2},
	\label{eq:4}
\end{aligned}
\end{equation}
where we have taken into account that $t_{\pi \bar K}$ and $t_{\pi K}$ are the same;

$b)$ $\eta$ scattering
\begin{equation}
\begin{aligned}
	t_1 &= t_{\eta K^{*}}^{\,I=1/2},\\[2mm]
	t_2 &= t_{\eta \bar K}^{\,I=1/2} = t_{\eta K}^{\,I=1/2}.
	\label{eq:3}
\end{aligned}
\end{equation}
The amplitudes $t_{\pi K^{*}}^{I=3/2}$, $t_{\pi K^{*}}^{I=1/2}$, $t_{\pi K}^{I=3/2}$, $t_{\pi K}^{I=1/2}$, $t_{\eta K^{*}}^{\,I=1/2}$ and $t_{\eta K}^{\,I=1/2}$ appearing in $t_1,\, t_2$ of Eqs.~\eqref{eq:4} and \eqref{eq:3} are evaluated in the Appendix.

One defines the partition matrices $\tilde{T}_{ij}$ corresponding to the diagrams of Fig.~\ref{Fig1}, as the sum of diagrams where the external particle interacts first with particle $i$ of the cluster and finishes with interaction with particle $j$. Then we define the matrix $\tilde{T}$ as
\begin{equation}
	\tilde{T}= \begin{pmatrix} \tilde{T}_{11} & \tilde{T}_{12} \\[1mm] \tilde{T}_{21} & \tilde{T}_{22} \end{pmatrix},
	\label{eq:5}
\end{equation}
where the elements $\tilde{T}_{ij}$ are obtained as
\begin{equation}
	\begin{split}
		& \tilde{T}_{11} = \dfrac{\tilde{t}_1}{1 - \tilde{t}_1 \,\tilde{t}_2 \,G_0^2}, ~~~~~~~~~~~~
		\tilde{T}_{22} = \dfrac{\tilde{t}_2}{1 - \tilde{t}_1\,\tilde{t}_2 \,G_0^2}, \\[2mm]
		& \tilde{T}_{12} = \tilde{T}_{21} = \dfrac{\tilde{t}_1 \,\tilde{t}_2 \, G_0}{1 - \tilde{t}_1 \,\tilde{t}_2  \,G_0^2},
	\end{split}
	\label{eq:6}
\end{equation}
and $G_0(\sqrt{s})$ stands for the propagator of the external particle modulated with the wave function of the cluster, given by
\begin{align}\label{eq:7}
	G_0(\sqrt{s}) &= \int \dfrac{\mathrm{d}^3 q}{(2\pi)^3} 
	\;\dfrac{F_c(q)}{\sqrt{s} - \omega_{ex}(\vec q\,) - \omega_c(\vec q \,) + i \epsilon} \;\dfrac{1}{2 \,\omega_{ex}(\vec q\,)} \nonumber \\[2mm]
	&\times \dfrac{1}{2 \,\omega_{c}(\vec q\,)}  \;\Theta\left(q_{\mathrm{max}}^{(1)} - q_1^*\right)  \Theta\left(q_{\mathrm{max}}^{(2)} - q_2^*\right),
\end{align}
with $\omega_{ex}(\vec q\,) = \sqrt{m_{ex}^2 + \vec{q}^{\;2}}$, where $m_{ex}$ is the mass of the external particle ($\pi^0$ or $\eta$), and $\omega_c(\vec q\,) = \sqrt{M_c^2 + \vec{q}^{\;2}}$. $F_c(q)$ is the form factor of the cluster, given by
\begin{equation}\label{eq:8}
	\begin{aligned}
		F_c(q) &= \dfrac{F(q)}{N}, \\[2mm]
		F(q) &= \int\limits_{\substack{|\vec{p}\,| < q_{\mathrm{max}} \\ |\vec{p} - \vec{q}\,| < q_{\mathrm{max}}}} 
		\dfrac{\mathrm{d}^3 p}{(2\pi)^3} \;
		\dfrac{1}{M_c - \omega_{K^*}(\vec{p}\,) - \omega_{\bar K}(\vec{p}\,)} \\
		&\quad \times \dfrac{1}{M_c - \omega_{K^*}(\vec{p} - \vec{q}\,) - 
			\omega_{\bar K}(\vec{p} -\vec{q}\,)}, \\[2mm]
		N &= F(0) \\
		&= \int\limits_{|\vec{p}\,| < q_{\mathrm{max}}} \dfrac{\mathrm{d}^3 p}{(2\pi)^3} \,
		\left[ \dfrac{1}{M_c - \omega_{K^*}(\vec{p}\,) - \omega_{\bar K}(\vec{p}\,)} \right]^2.
	\end{aligned}
\end{equation}
The magnitude $q_{\mathrm{max}}$ in Eq.~\eqref{eq:8} is the cutoff regularizing the $G$ loops of the Bethe-Salpeter series in $t = (1-VG)^{-1}V$, with $V$ the potential, when one studies the $K^* \bar{K}$ interaction that generates the $f_1(1285)$ molecule \cite{Roca:2010tf}, which renders the wave function of the type 
\begin{equation}\label{eq:9}
	\Psi(p)\sim \frac{\Theta\left(q_{\mathrm{max}}-|\vec{p}\,|\right)}
	{M_c - \omega_{K^*}(\vec{p}\,) - \omega_{\bar K}(\vec p\,)}.
\end{equation}
The limits of $\vec{p}$ in the integral of Eq.~\eqref{eq:8} make the integral convergent, and $F(q)$ vanishes for $|\vec{q}\,| > 2\, q_{\mathrm{max}}$.
In Eq.~\eqref{eq:7}, the values of $q^{(1)}_{\mathrm{max}}$, $q^{(2)}_{\mathrm{max}}$ are the cutoffs used to regularize the $\pi^0(\eta) K^*$ and $\pi^0(\eta) \bar{K}$ loops when constructing the corresponding scattering matrices, $t_1,\,t_2$.
On the other hand, $q^*_1,\,q^*_2$ are the $\pi^0(\eta)$ momenta in the rest frames of $\pi^0(\eta) K^*$ and $\pi^0(\eta) \bar{K}$ respectively, obtained assuming the momentum transferred to be shared between the initial and final particles of the cluster \cite{Boffi:1991nh, Carrasco:1991we}. One has
\begin{equation}\label{eq:10}
	\vec{q}^{\,*}_i = \vec{q}\,\left( 1 - \frac{1}{2}\frac{m_{ex}}{m_{ex} + m_{i}} \right),
\end{equation}
with $m_1=m_{K^*}$, $m_2=m_{\bar K}$.

\begin{figure}[t]
	\begin{center}
		\includegraphics[width=0.48\textwidth]{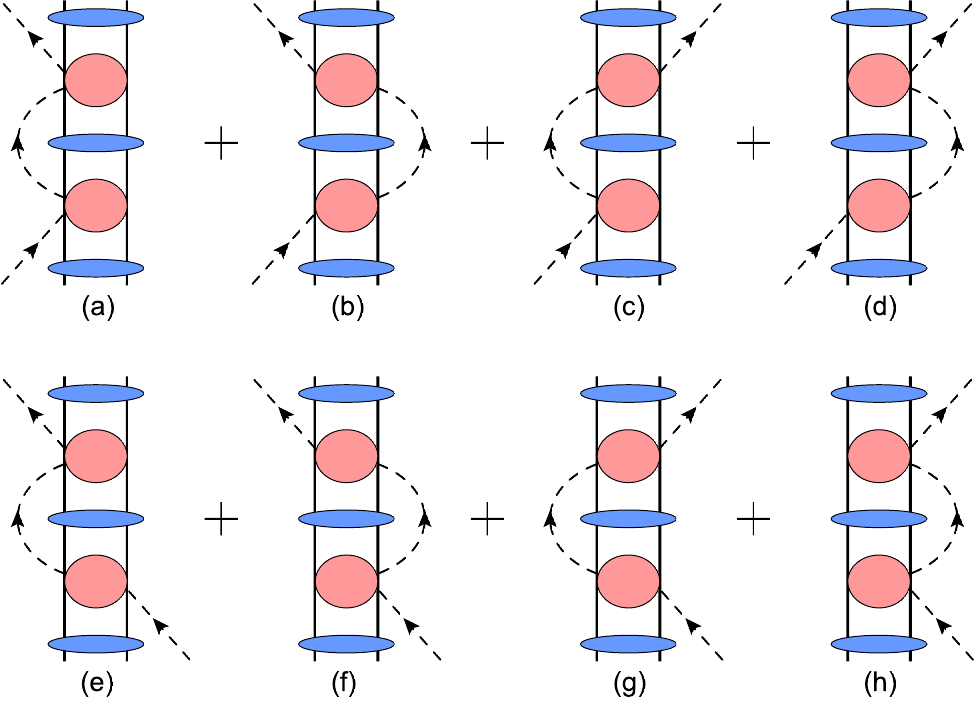}
	\end{center}
	\vspace{-0.5cm}
	\caption{Diagrams considering the elastic propagation of the external particle and the cluster $f_1(1285)$ as a whole.}
	\label{Fig2}
\end{figure}

The elastic unitarity in the $\pi^0(\eta) f_1(1285)$ is obtained by summing the diagrams of Fig.~\ref{Fig2}.
This gives rise to the matrix $\tilde{T}'_{ij}$, where, again, one sums over all diagrams in which the first scattering is with particle $i$ and the final one with particle $j$ of the cluster. 
One defines the $\tilde{T}'$ matrix as
\begin{equation}\label{eq:11}
	\tilde{T}' = \begin{pmatrix} \tilde{T}'_{11} & \tilde{T}'_{12} \\[1.5mm]
		\tilde{T}'_{21} & \tilde{T}'_{22} \end{pmatrix},
\end{equation}
and obtains $\tilde{T}'$ via a Bethe-Salpeter equation in which $\tilde{T}_{ij}$ plays the role of an optical potential, via
\begin{equation}\label{eq:12}
	\tilde{T}'=\left[ 1-\tilde{T}\,G_c \right]^{-1}\;\tilde{T},
\end{equation}
where
\begin{equation}\label{eq:13}
	G_c=\begin{pmatrix}
		G_c^{(1)} & 0 \\[1mm]
		0 & G_c^{(2)}
	\end{pmatrix},
\end{equation}
and
\begin{align}\label{eq:14}
		G_c^{(i)}(\sqrt{s}) &= \int \dfrac{\mathrm{d}^3 q}{(2\pi)^3}\;
		\dfrac{\left[ F_c^{(i)}(q) \right]^2}{\sqrt{s} - \omega_{ex}(\vec q\,) - \omega_c(\vec q\,) + i\epsilon} \nonumber \\[2mm]
		&\quad \times \dfrac{1}{2\,\omega_{ex}(\vec q\,)} \;\dfrac{1}{2 \, \omega_{c}(\vec q\,)} \;\Theta\left(q_{\mathrm{max}}^{(i)} - q_i^*\right),
\end{align}
with \cite{Yamagata-Sekihara:2010kpd}
\begin{equation}
\begin{aligned}\label{eq:15}
		F_c^{(1)}(\vec{q}\,) &= F_c\left( \frac{m_{\bar K}}{m_{K^*}+m_{\bar K}} \vec{q}\,\right), \\[2mm]
		F_c^{(2)}(\vec{q}\,) &= F_c\left( \frac{m_{K^*}}{m_{K^*}+m_{\bar K}} \vec{q}\,\right).
\end{aligned}
\end{equation}
The $t_1, t_2$ amplitudes depend on $\sqrt{s_1}, \sqrt{s_2}$ invariant masses respectively which are calculated assuming the binding of the cluster to be absorbed by each particle of the cluster proportional to their mass, as
\begin{align}\label{eq:16}
		s_1(\pi^0(\eta) K^*) &= \left(p_{ex} + p_{K^*}\right)^2 \\[1mm]
		&= m_{ex}^2 + \left(\xi  \,m_{K^*}\right)^2 + 2\,\xi \,m_{K^*}\,q^0,\nonumber\\[2mm]
		s_2(\pi^0(\eta) \bar K) &= \left(p_{ex} + p_{\bar K}\right)^2 \nonumber\\[1mm]
		&= m_{ex}^2 + \left(\xi  \,m_{\bar K}\right)^2 + 2\,\xi \,m_{\bar K}\,q^0, \label{eq:16n}
\end{align}
with $q^0$ being the $\pi^0 (\eta) $ energy in the cluster rest frame
\begin{equation}\label{eq:17}
	q^0=\frac{s-m_{ex}^2-M_c^2}{2 \,M_c},
\end{equation}
and
\begin{equation}\label{eq:18}
	\xi=\frac{M_c}{m_{K^*} + m_{\bar K}}.
\end{equation}

The final amplitude for the $\pi^0 (\eta) f_1$ scattering is given by \cite{Agatao:2025ckp}
\begin{align}\label{eq:19}
	T^{\,\mathrm{tot}}&=\sum_{i,j} \tilde{T}'_{ij}\nonumber\\[2mm]
	&=\frac{
		\tilde{t}_1+\tilde{t}_2+\left(2 \, G_0-G_c^{(1)}-G_c^{(2)}\right) \tilde{t}_1  \, \tilde{t}_2}
	{1-G_c^{(1)} \, \tilde{t}_1-G_c^{(2)} \,\tilde{t}_2-\left(G_0^2-G_c^{(1)} \, G_c^{(2)}\right) \tilde{t}_1 \,\tilde{t}_2},
\end{align}
which allows for an easy proof of the elastic unitarity \cite{Agatao:2025ckp}.

\subsection{Scattering length and effective range}
Using the relationship of our amplitude to the standard one in Quantum Mechanics, $f^{\mathrm{QM}}$, we have
\begin{align}\label{eq:20}
		-8\pi\sqrt{s} \left(T^{\,\mathrm{tot}}\right)^{-1} &= (f^{\mathrm{QM}})^{-1} \nonumber \\[1mm]
		&\simeq  -\dfrac{1}{a} + \dfrac{1}{2}\,r_0 \,q_{\mathrm{cm}}^2 - i q_{\mathrm{cm}},
\end{align}
where $q_{\mathrm{cm}}$ is the $\pi^0(\eta)$ momentum in the $\pi^0(\eta) f_1$ rest frame. Then, by looking at the amplitude around the threshold, it follows immediately that
\begin{align}
	a &= \dfrac{T^{\,\mathrm{tot}}}{8\,\pi\sqrt{s}}\,\Big|_{\mathrm{th}}, \label{eq:21} \\[2mm]
	r_0 &= \dfrac{1}{\mu} \left[ \dfrac{\partial}{\partial \sqrt{s}} \left( -8\,\pi\sqrt{s} \,\left(T^{\,\mathrm{tot}}\right)^{-1} + iq_{\mathrm{cm}} \right) \right]_{\mathrm{th}},\label{eq:22}
\end{align}
where $\mu$ is the reduced mass of the $\pi^0(\eta) f_1$ system.

\subsection{Correlation function}
The $\pi^0(\eta) f_1$ correlation function is then given by
\begin{align}
		C_{\pi^0(\eta)f_1}(p) = & 1 + 4\pi \int_{0}^{\infty} \mathrm{d}r \, r^2 \,S_{12}(r) \nonumber \\[1.5mm]
		& ~~~~~\times \left\{ \left|j_0(pr) + TG\right|^2 - j_0^2(pr) \right\},
	\label{eq:23}
\end{align}
where
\begin{equation}
	TG=\left(
	\tilde{T}'_{11}+\tilde{T}'_{21}
	\right)G_1(\sqrt{s},r)+\left(
	\tilde{T}'_{12}+\tilde{T}'_{22}
	\right)G_2(\sqrt{s},r),
	\label{eq:24}
\end{equation}
with
\begin{equation}
	S_{12}(r)=\frac{1}{\left(
		4\pi R^2
		\right)^{3/2}}\;e^{-r^2/4R^2},
	\label{eq:25}
\end{equation}
with $R$ the radius of the source, and $G_1,\,G_2$ given by
\begin{align}
		G_i(\sqrt{s},r) &= \int \dfrac{\mathrm{d}^3 q}{(2\pi)^3} \;
		\dfrac{ j_0(qr)\,F_c^{(i)}(q) }{\sqrt{s} - \omega_{ex}(\vec q\,) - \omega_c(\vec q\,) + i\epsilon} \nonumber \\[2mm]
		&\quad \times \dfrac{1}{2\,\omega_{ex}(\vec q\,)} \;\dfrac{1}{2\,\omega_{c}(\vec q\,)}\;\Theta\left(q_{\mathrm{max}}^{(i)} - q_i^*\right).
	\label{eq:26}
\end{align}

\section{Results}\label{sec:res}
The first results that we present are for the scattering length $a$ and the effective range $r_0$. We obtain:

$a)$ for the $\pi^0 f_1$ system
\begin{align}
	a &= -0.05\,\mathrm{fm}, \label{eq:27} \\[2mm]
	r_0 &= (-47.37-i\, 0.49)\,\mathrm{fm}. \label{eq:28}
\end{align}
As one can see, this is a special case with a very small scattering length and a large effective range.

$b)$ for the $\eta f_1$ system
\begin{align}
	a &= (0.29-i\, 0.08)\,\mathrm{fm}, \label{eq:29} \\[2mm]
	r_0 &= (0.47+i\, 0.50)\,\mathrm{fm}. \label{eq:30}
\end{align}
The value of $a$ in this latter case is about half the one of the $K f_1$ system and $r_0$ has also about half the size of the $K f_1$ system and with opposite sign.

\begin{figure}[t]
	\begin{center}
		\includegraphics[width=0.48\textwidth]{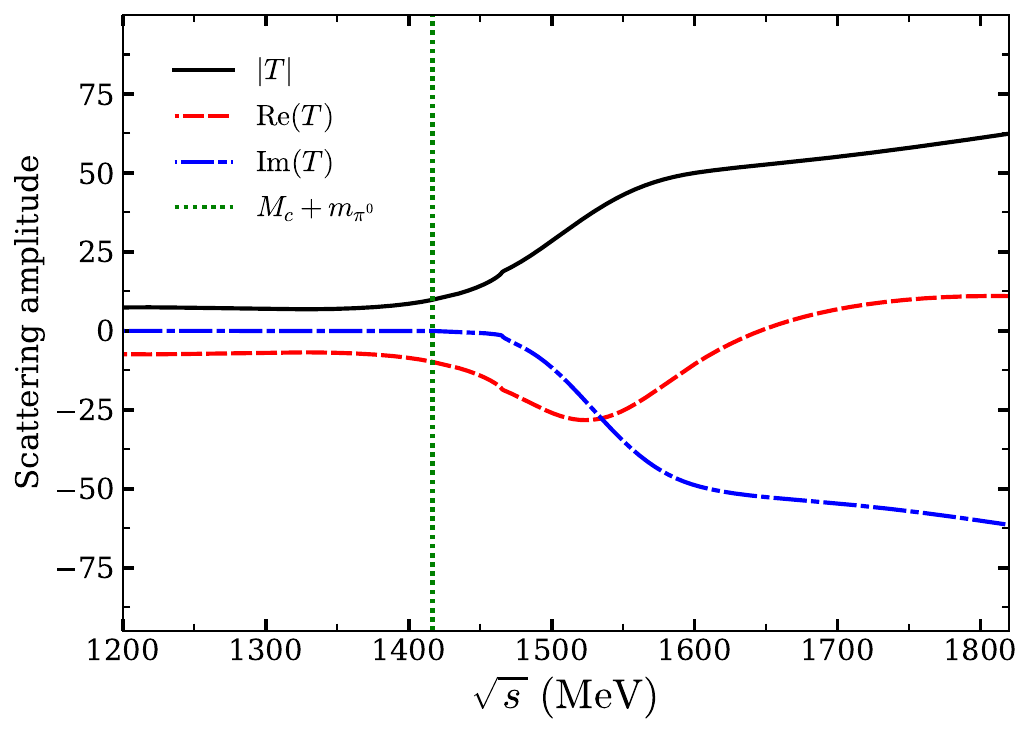}
	\end{center}
	\vspace{-0.5cm}
	\caption{$\pi^0 f_1(1285)$ scattering amplitude $T_{\pi^0 f_1}^{\,\mathrm{tot}}$ as a function of $\sqrt{s}$.}
	\label{Fig3}
\end{figure}
\begin{figure}[t]
	\begin{center}
		\includegraphics[width=0.48\textwidth]{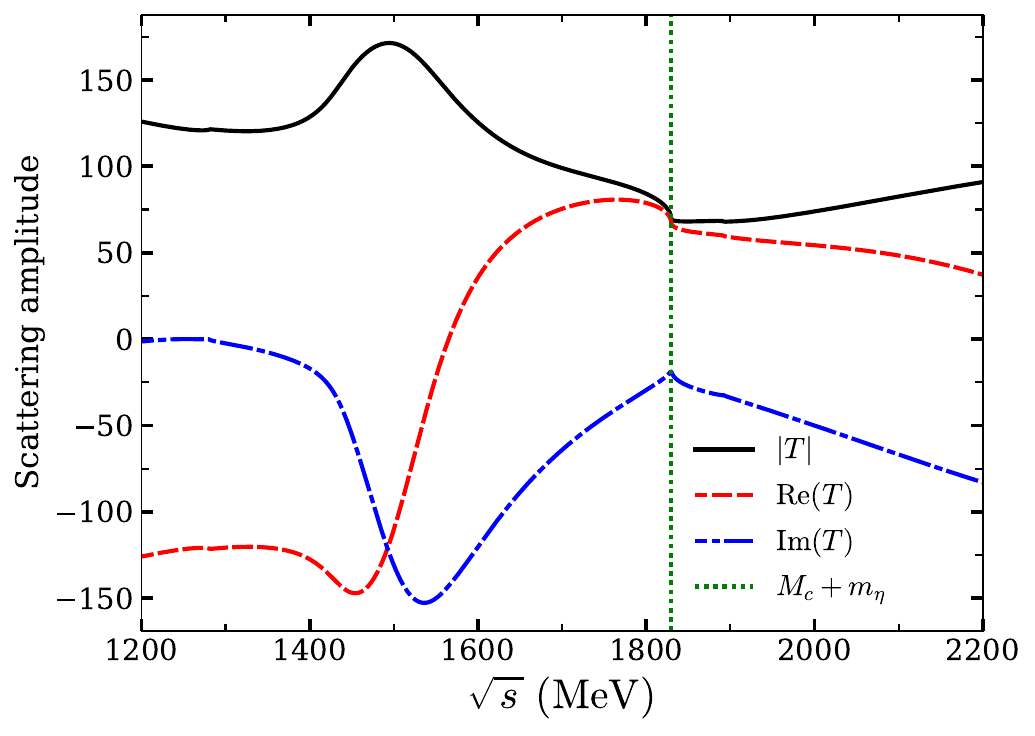}
	\end{center}  
	\vspace{-0.5cm}
	\caption{$\eta f_1(1285)$ scattering amplitude $T_{\eta f_1}^{\,\mathrm{tot}}$ as a function of $\sqrt{s}$.}
	\label{Fig4}
\end{figure}

Next we show the results for the total amplitude $\pi^0 f_1(1285)$ as a function of $\sqrt{s}$ in Fig.~\ref{Fig3}.
In Fig.~\ref{Fig4} we show the results for $\eta f_1(1285)$ scattering.

In Fig.~\ref{Fig3}, for the $\pi^0 f_1(1285)$ amplitude, we observe a smooth behaviour at threshold and below, but around $1500-1600\mev$ we observe a curious structure. 
The amplitude behaves like a resonance with a phase of $\pi/2$, where the roles of the real and imaginary parts are interchanged.
The real part has a bump and the imaginary part is like the real part of a Breit-Wigner resonance, adding to it a constant. 
It is unclear whether this structure is related to the $\pi_1(1400-1600)$ state reported in the PDG, but it is intriguing to see that a structure in the amplitude emerges precisely at the energy where the $\pi_1(1400-1600)$ is claimed experimentally.

\begin{figure}[t]
	\begin{center}
		\includegraphics[width=0.48\textwidth]{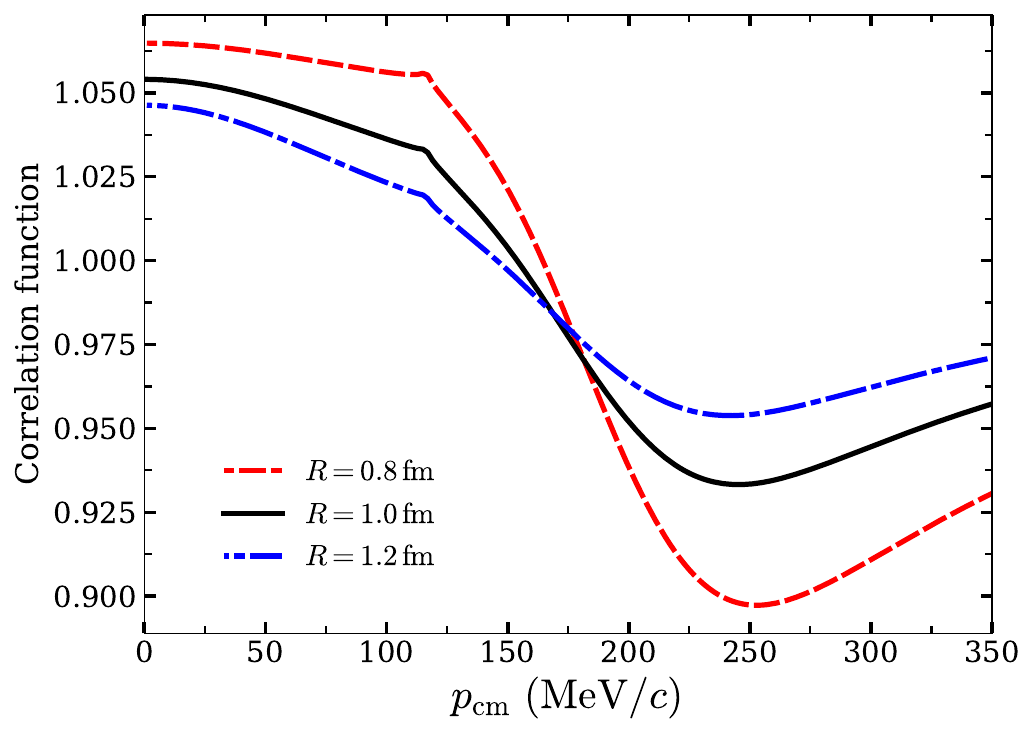}
	\end{center}  
	\vspace{-0.5cm}
	\caption{Correlation functions for the $\pi^0 f_1(1285)$ system as a function of momentum $p_{\rm cm}$, with different values of the source radius $R$.}
	\label{Fig5}
\end{figure}
\begin{figure}[t]
	\begin{center}
		\includegraphics[width=0.48\textwidth]{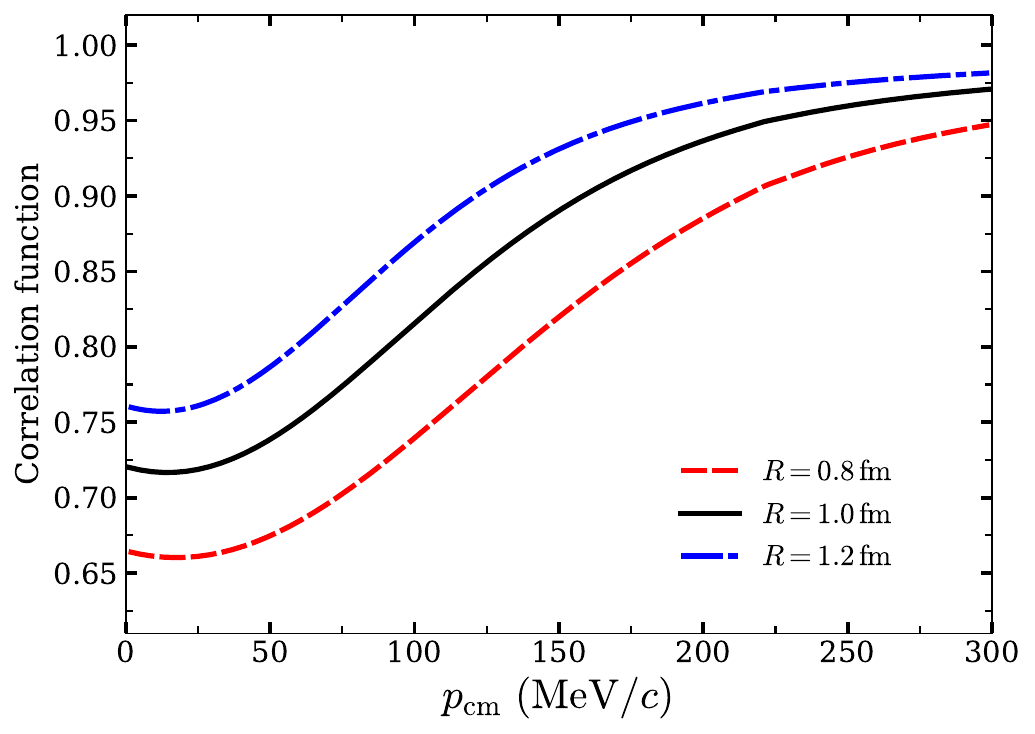}
	\end{center}  
	\vspace{-0.5cm}
	\caption{Correlation functions for the $\eta f_1(1285)$ system as a function of momentum $p_{\rm cm}$, with different values of the source radius $R$.}
	\label{Fig6}
\end{figure}

Concerning the $\eta f_1$ amplitude, the first comment is that we do not see any structure around $1855\mev$ that could be associated with the $\eta_1(1855)$. 
We can only state that we get a strong cusp structure around the threshold at $1830\mev$.
The second comment is that we see some structure around $1500\mev$. 
This is, however, about $330\mev$ below the $\eta f_1(1285)$ threshold, where we are reluctant to trust our results.
We can reasonably trust our results around $150\mev$ above and below threshold (the range of the correlation functions energies in Ref.~\cite{Lauraser} and in Fig.~\ref{Fig6} below). Thus, we prefer not to make any claim concerning this structure.

Next we show the results for the correlation functions. 
In Fig.~\ref{Fig5} we show the results for the $\pi^0 f_1(1285)$ correlation function.
In Fig.~\ref{Fig6} we show the same results for the $\eta f_1(1285)$ correlation function.
For the case of $\pi^0 f_1$, we see a correlation function with values very close to $1$, indicating a weak interaction.
The value of the correlation function decreases with increasing momentum at small values of $p_{\rm cm}$.
The kink appearing around $p_{\rm cm} \simeq  110\mev$ corresponds to the opening of the $\pi K$ threshold in the $t_2$ amplitude, as one can see by applying Eq.~\eqref{eq:16n} that relates $s$ to $s_2$. 
This corresponds to $\sqrt{s} \simeq 1465\mev$, and is also barely seen in Fig.~\ref{Fig3} at this energy. 
There are no kinks related to the $\pi K^*$ threshold because we use the convolved $G$ functions for the $K^*$. 
In the case of the $\eta f_1$ interaction, the kink expected for the $\eta K$ threshold is small and not seen in the figures. 
The trend of the correlation function is similar to the one obtained for the $K f_1(1285)$ case \cite{Jia:2026dpl} but closer to unity, indicating a weaker interaction than for the $K f_1(1285)$ case. 
The shape obtained is similar to that of the $p f_1(1285)$ observed in the ALICE experiment \cite{Lauraser}, however, the strength is much weaker (with the correlation function closer to unity) than that observed in the $p f_1(1285)$ experiment.

\section{Discussion and conclusions}\label{sec:con} 
We have studied the interaction of a $\pi^0$ (or $\eta$) with the $f_1(1285)$ resonance, assuming that the $f_1(1285)$ is a $K^* \bar K-\bar{K^*}K$ bound state. 
We have used a framework to calculate it as one does in calculations of an external particle with a nucleus, defining an optical potential and later on constructing the Lippmann-Schwinger equation from this optical potential. 
In our case, the role of the nucleus is played by the two-particle cluster, the $f_1(1285)$, while the external particle is either a $\pi^0$ or an $\eta$. 
The optical potential is obtained with the formalism of the fixed center approximation (FCA) to the three-body Faddeev equations, where the main assumption is that the two-particle cluster remains intact during the interaction, a situation perfectly suited to the present case, where we have the $f_1(1285)$ particle at both the beginning and the end of the interaction. 
 
The study of this interaction becomes more relevant when compared to other calculations where the kernel of the $\pi^0 (\eta)\, f_1(1285)$ interaction is obtained by assuming the $f_1(1285)$ as a matter source and using the Weinberg-Tomozawa interaction, which happens to vanish in these two cases. 
We have seen that in our case we go beyond the $O(1/f^2)$ level of the Weinberg-Tomozawa interaction, because our total amplitudes use as input individual $t_i$ matrices that contain higher order terms than $O(1/f^2)$, obtained from coupled-channel calculations implying unitarity in coupled channels, and these amplitudes do not vanish. 
  
We have calculated the total amplitudes for the $\pi^0 (\eta)\, f_1(1285)$ interaction and find that we do not obtain clear signals for the $\pi_1(1400)$ or $\pi_1(1600)$, nor for the $\eta_1(1855)$ resonances from our approach. 
However, we note the appearance of a weak structure around $1500-1600\mev$ in the $\pi^0 f_1(1285)$ amplitude, and a strong cusp structure in the $\eta f_1(1285)$ amplitude at the $\eta f_1(1285)$ threshold around $1830\mev$.

We should mention that in Ref.~\cite{Yan:2023vbh}, in spite of having the $\pi^0 (\eta)\, f_1(1285)$ interaction null, the use of coupled channels involving other pseudoscalar meson-axial vector meson channels leads finally to the $\pi_1(1600)$ and $\eta_1(1855)$ states. 
In our case one could in principle do that; however, rather than dealing with the overlap of two particles in the framework of Ref.~\cite{Yan:2023vbh}, we would have to deal with the overlap of three particles in this hypothetical calculation, since the axial-vector mesons are molecular states in our approach, and we deem this three-body overlap to be more difficult than in the two-body case. Thus, we do not anticipate a relevant effect from this hypothetical calculation. 
This renders the theoretical explanation of the $\pi_1(1600)$ and $\eta_1(1855)$ states a bit further than previously assumed. 
   
As to the work of Ref.~\cite{Zhang:2016bmy}, the authors claim to get the $\pi_1(1600)$ from the $\pi f_1(1285)$ interaction, using the FCA.
It is unclear whether the structure that we find for the $\pi^0 f_1(1285)$ amplitude around $1500-1600\mev$ is related to the state found in Ref.~\cite{Zhang:2016bmy}. 
There are some differences between our approach and that of Ref.~\cite{Zhang:2016bmy}, since we use the new formalism implementing elastic unitarity in the FCA amplitude. 
As previously discussed, the standard FCA amplitude does not satisfy unitarity at threshold; hence cannot be reliably used near threshold, and becomes more uncertain when one goes further above it. 
This problem was already anticipated, when comparing the FCA results with Faddeev calculation for a state appearing above threshold, the $\phi(2170)$, where one could see the failure of the FCA in that case \cite{MartinezTorres:2010ax}.

In the present work we have evaluated the scattering length and effective range for the  $\pi^0 (\eta)\, f_1(1285)$ interaction, as well as the corresponding correlation functions. 
With the advent of the first experimental measurement of the correlation function for a proton and the $f_1(1285)$, a door has been opened for future studies of the interaction of other particles with the $f_1(1285)$ and with other resonances. 
It will be interesting to compare these predictions with the experimental results. 
The good agreement between the predictions obtained using the present formalism for the interaction of a proton and the $f_1(1285)$ in Ref.~\cite{Encarnacion:2026zas} and posterior experimental results \cite{Lauraser} makes us optimistic about the result of these comparisons.

\section{Uncertainties}\label{sec:uncertainties} 
As shown in Appendix A, the needed amplitudes are obtained using the chiral unitary approach in coupled channels, with the transition potential obtained from chiral Lagrangians at lowest order. Possible contributions from higher order potentials are empirically accommodated by the choice of $q_{\mathrm{max}}$ to fit scattering data or the mass of the generated resonances. In one channel, for simplicity of the argumentation, one has $t=[V^{-1}-G]^{-1}$, hence, changes in $V$ can be accommodated by changing the $G$ function via the cutoff, $q_{\mathrm{max}}$, in the loop integration.
We used $q_{\mathrm{max}}=600\mev$ in the evaluation of the relevant scattering matrices in Appendix A, and we would like to see how stable the results can be by reasonable changes of $q_{\mathrm{max}}$.

What we say for the scattering matrices $t_1$, $t_2$, can equally be said for the wave function of the cluster in Eq.~\eqref{eq:9}.
In Ref.~\cite{Roca:2005nm} the value of $q_{\mathrm{max}}$ used to generate the different axial vector resonances was $q_{\mathrm{max}} \simeq 1000\mev$, and the mass of the $f_1(1285)$ was predicted quite close to the physical one. Here, by tuning $q_{\mathrm{max}}$ to get the exact mass of the $f_1(1285)$, we obtain $q_{\mathrm{max}}=971.5\mev$. We then check the stability of the results by changing $q_{\mathrm{max}}$ by $10\%$ up and down. The changes in the scattering length and effective range of $\pi^0 f_1(1285)$ are much smaller than $1\%$.
The changes are bigger for the $\eta f_1(1285)$ system: around $2\%$ for the scattering length and as much as $15\%$ for the effective range. The changes in the correlation functions for both systems are very small, much smaller than those induced by the change of $q_{\mathrm{max}}$ in the evaluation of $t_1$ and $t_2$, which we show below. 

The large value of $r_0$ in the $\pi^0 f_1(1285)$ system of about 47 fm seems abnormally large. We could think that this could change if we take into account the
width of the $f_1(1285)$, which has been ignored so far. We can take into account this width by performing a convolution of the results with the spectral function of the $f_1(1285)$ resonance $S(M_{\mathrm{inv}}) = - \frac{1}{\pi} \mathrm{Im} \frac{1}{M_{\mathrm{inv}}^2 - M_{f_1}^2 + i M_{f_1} \Gamma_{f_1}}$, as done in the study of the $K f_1(1285)$ system \cite{Jia:2026dpl}. The changes in $a$ and $r_0$ for the $\pi^0 f_1(1285)$ system are small.
The new values with the convolution are
\begin{align}
	a &= -0.05\,\mathrm{fm}, \label{eq:32} \\[2mm]
	r_0 &= (-50.38-i\, 0.34)\,\mathrm{fm}. \label{eq:33}
\end{align}
Thus we see no change in $a$ and a $6\%$ change in $r_0$. We can see that the convolution of the $f_1(1285)$ width has not changed the large value of the effective range. 
In order to understand the large value of $r_0$, one can go to Eq.~\eqref{eq:20} and Fig.~\ref{Fig3}. 
One can find numerically that the structure found above threshold demands the small value of $a$ and the large value of $r_0$ to be reproduced.

\begin{figure}[t]
	\begin{center}
		\includegraphics[width=0.48\textwidth]{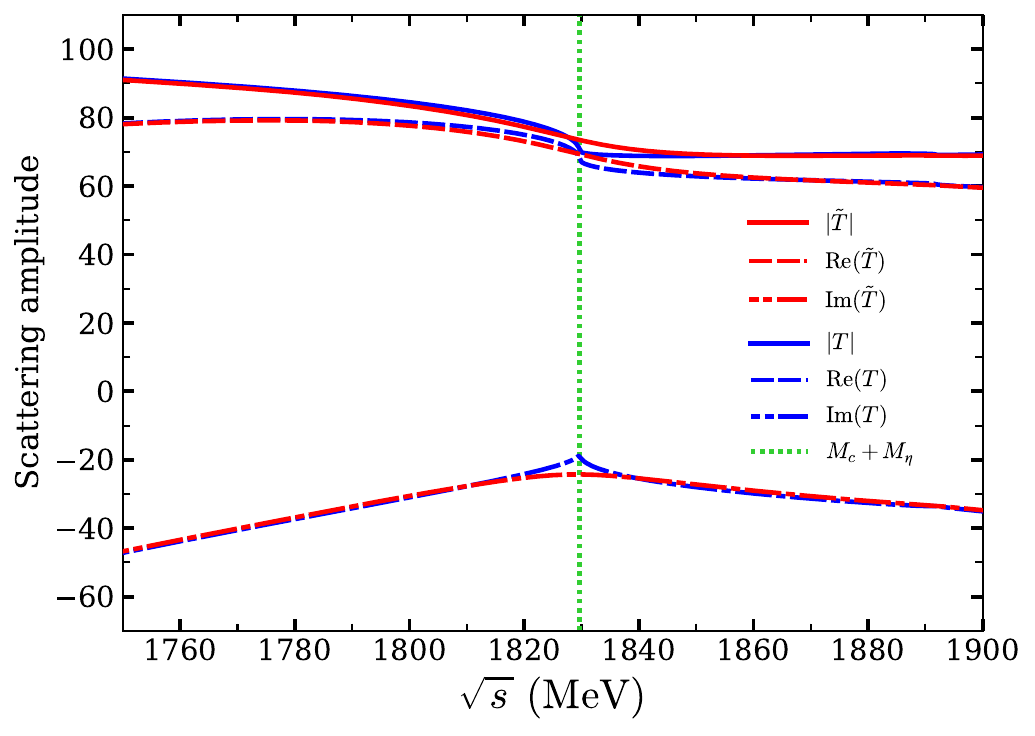}
	\end{center}  
	\vspace{-0.5cm}
	\caption{Changes in the scattering amplitude induced by the convolution due to the $f_1(1285)$ width.}
	\label{Fig7}
\end{figure}
\begin{figure}[t]
	\begin{center}
		\includegraphics[width=0.48\textwidth]{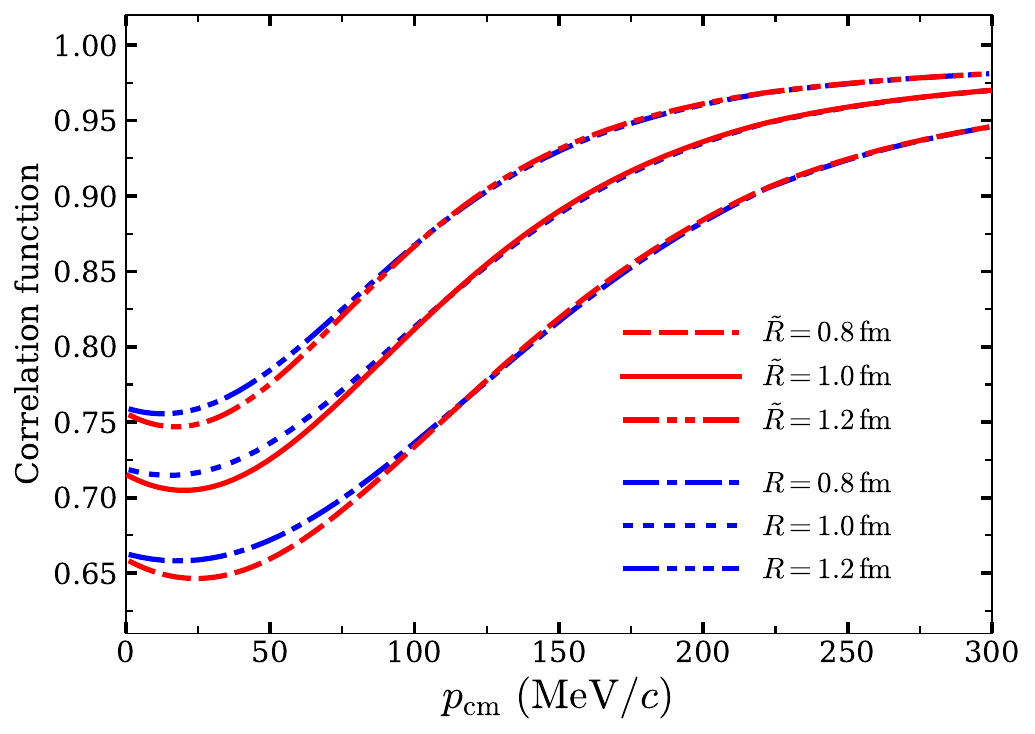}
	\end{center}  
	\vspace{-0.5cm}
	\caption{Changes in the correlation function induced by the convolution due to the $f_1(1285)$ width.}
	\label{Fig8}
\end{figure}

We also see that the changes in the amplitude and the correlation function of $\pi^0 f_1(1285)$ induced by the convolution are small, much smaller than those generated by the changes of $q_{\mathrm{max}}$ in $t_1$, $t_2$ that we show below.

The changes in the $\eta f_1(1285)$ observables are larger than before. We obtain now
\begin{align}
	a &= (0.30-i\, 0.10)\,\mathrm{fm}, \label{eq:34} \\[2mm]
	r_0 &= (-1.46+i\, 0.57)\,\mathrm{fm}. \label{eq:35}
\end{align}
Compared with the results of Eqs.~\eqref{eq:29} and \eqref{eq:30}, we see that the changes in $a$ are moderate, however, the value of $r_0$ has changed drastically and even changes sign. It is also worth showing how the scattering amplitude and the correlation function have changed.
We show that in Fig.~\ref{Fig7} for the three-body amplitude and in Fig.~\ref{Fig8} for the correlation function. The changes are moderate, but one observes a softening of the amplitude at threshold and small changes of the correlation function at very small values of $p_{\mathrm{cm}}$.

\begin{figure}[t]
	\begin{center}
		\includegraphics[width=0.48\textwidth]{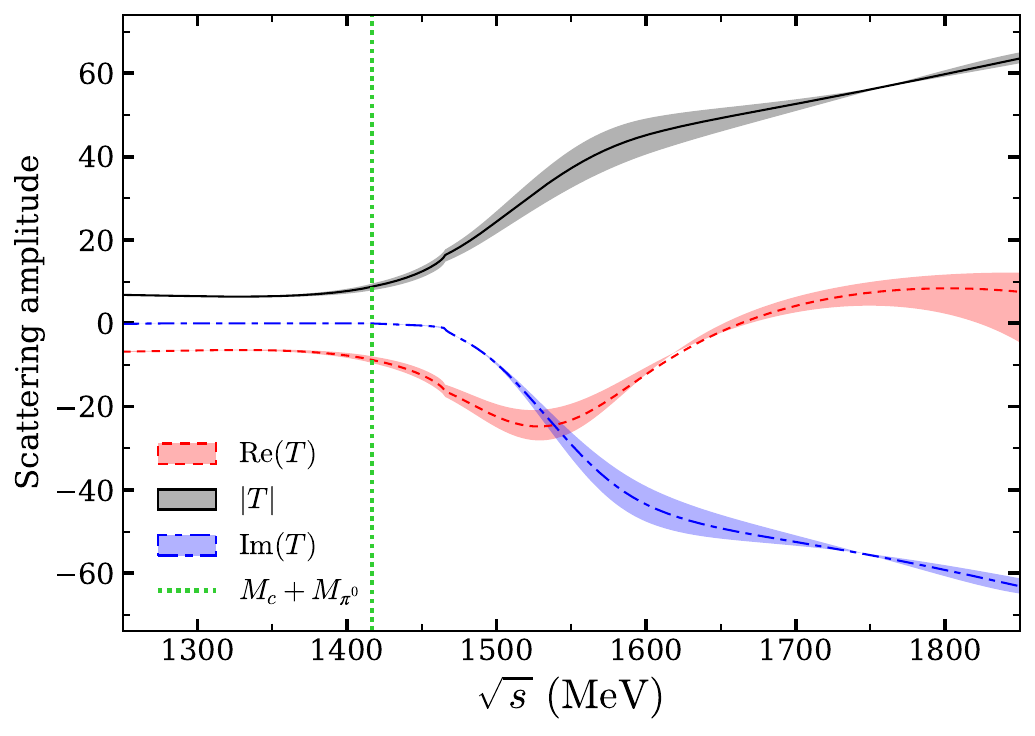}
	\end{center}  
	\vspace{-0.5cm}
	\caption{Band of values for the $\pi^0 f_1(1285)$ three-body amplitudes from uncertainties in the $\pi K$ amplitude.}
	\label{Fig100}
\end{figure}
\begin{figure}[t]
	\begin{center}
		\includegraphics[width=0.48\textwidth]{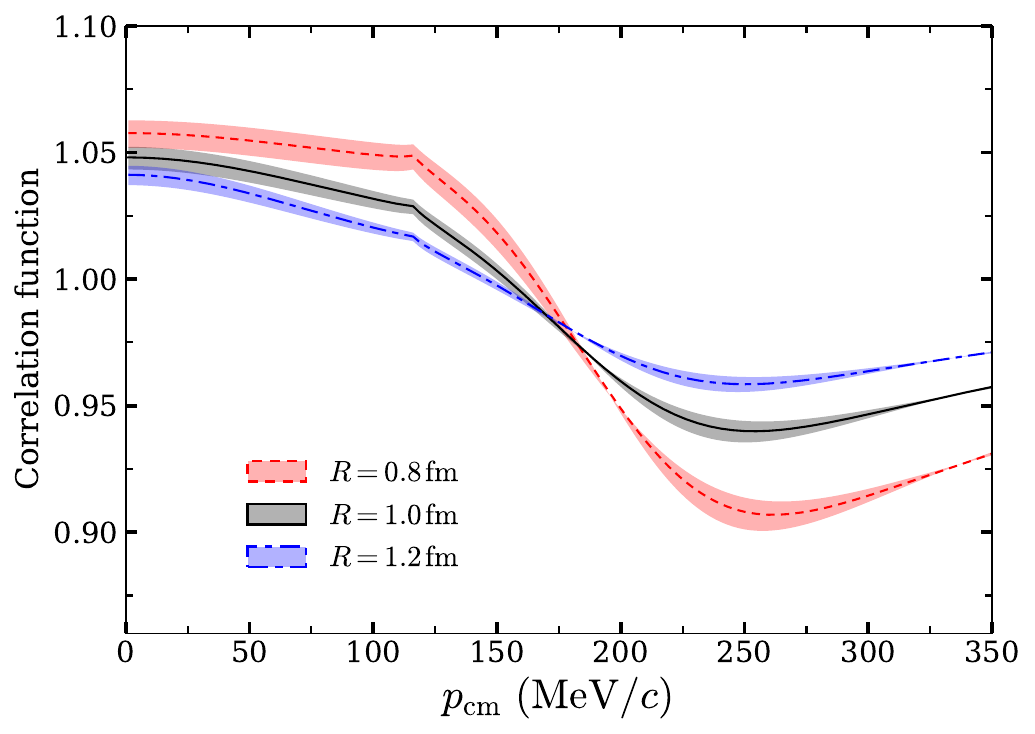}
	\end{center}  
	\vspace{-0.5cm}
	\caption{Band of values for the $\pi^0 f_1(1285)$ correlation function from uncertainties in the $\pi K$ amplitude.}
	\label{Fig101}
\end{figure}
\begin{figure}[t]
	\begin{center}
		\includegraphics[width=0.48\textwidth]{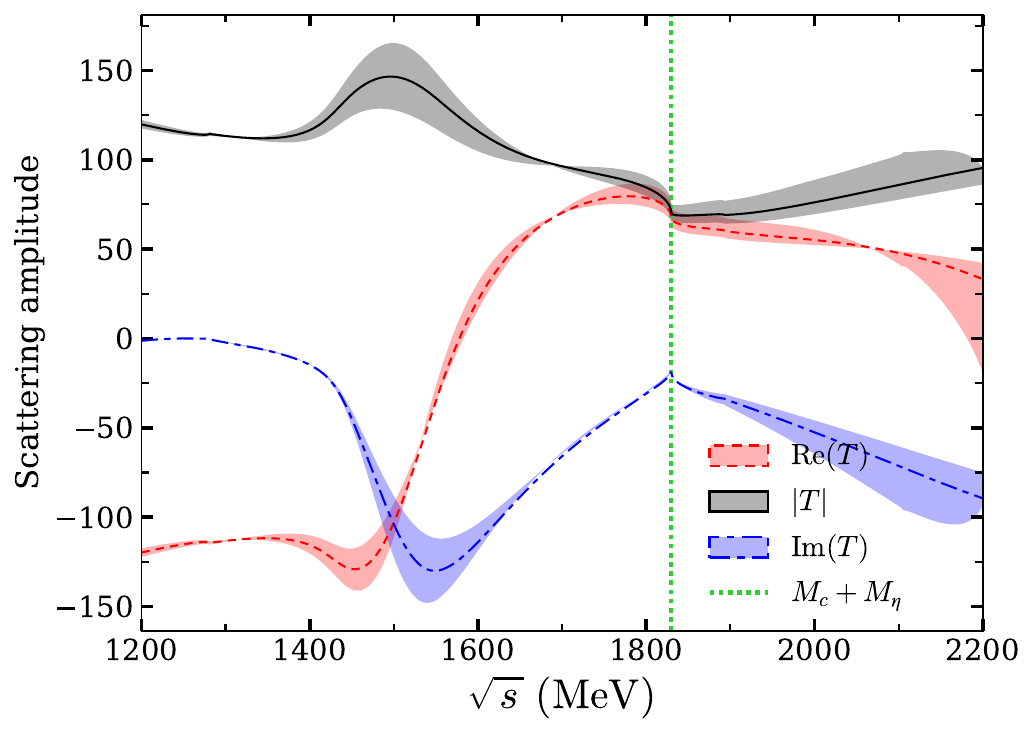}
	\end{center}  
	\vspace{-0.5cm}
	\caption{Band of values for the $\eta f_1(1285)$ amplitude from uncertainties in the $\pi K$ amplitude.}
	\label{Fig102}
\end{figure}
\begin{figure}[t]
	\begin{center}
		\includegraphics[width=0.48\textwidth]{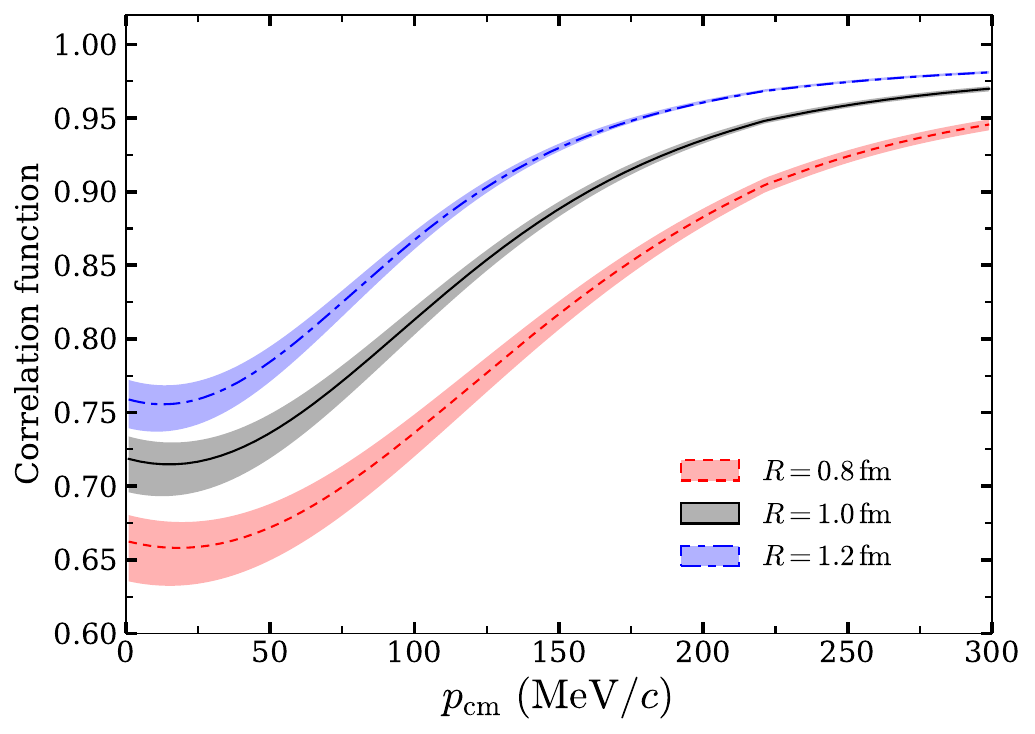}
	\end{center}  
	\vspace{-0.5cm}
	\caption{Band of values for the $\eta f_1(1285)$ correlation function from uncertainties in the $\pi K$ amplitude.}
	\label{Fig103}
\end{figure}

The biggest changes, and still moderate, stem from the change of $q_{\mathrm{max}}$ in the $\pi K$ amplitudes. 
By following Ref.~\cite{Toledo:2020zxj}, we assume that we can make a change of $10\%$ in the $\pi K$ potential which can be compensated by a change of $q_{\mathrm{max}}$ ($q_{\mathrm{max}}=600\mev$ to start with) by also $10\%$ up and down in order to find the $K_0^*(700)$ position at the same place. 
We find now a band of values
\begin{itemize}
	\item [i)] $\pi^0 f_1(1285)$:
	\begin{equation}
	\begin{aligned}
    & a \in [-0.04, -0.05] \,\text{fm}, \\
    & r_0 \in [-45.63 - i 0.62, -47.14 - i 0.41] \,\text{fm};
    \end{aligned}
	\end{equation}
	\item [ii)] $\eta f_1(1285)$:
	\begin{equation}
	\begin{aligned}
	a &\in [0.27-i0.08,\,0.32-i0.07]\,\mathrm{fm}, \\[1mm]
	r_0 &\in [0.37 + i\, 0.47,\,0.65 + i\, 0.53]\,\mathrm{fm}. 
\end{aligned}
\end{equation}
\end{itemize}
In Figs.~\ref{Fig100}-\ref{Fig103}, we show the band of values that stem for the three-body amplitude and the correlation function for the $\pi^0 f_1(1285)$ and $\eta f_1(1285)$ systems. 
The bands show bigger uncertainties than from the former discussed sources, but are still moderate, not changing the basic structure and characteristics of the calculated observables.

\section*{Acknowledgments}
This work is partly supported by the National Natural Science Foundation of China (NSFC) under Grants No. 12575081 and No. 12365019,
and by the Natural Science Foundation of Guangxi province under Grant No. 2023JJA110076,
and by the Central Government Guidance Funds for Local Scientific and Technological Development, China (No. Guike ZY22096024). 
This work is also partly supported by the National Key R{\&}D Program of China (Grant No. 2024YFE0105200).
J. Song acknowledges support from the
National Natural Science Foundation of China (NSFC)
under Grants No. 12405089 and No. 12247108, and by the China Postdoctoral Science Foundation under Grant No. 2022M720360 and No. 2022M720359, and by the Hainan Provincial Excellent Talent Team under the ``Four Talents'' Gathering Program of Hainan Province. 
This work is also partly supported by the Spanish Ministerio de Economia y Competitividad (MINECO) and European FEDER funds under Contracts No. FIS2017-84038-C2-1-PB, PID2020-112777GB-I00, and by Generalitat Valenciana under contract PROMETEO/2020/023. 
This project has received funding from the European Union Horizon 2020 research and innovation program under the program H2020-INFRAIA-2018-1, grant agreement No. 824093 of the STRONG-2020 project.

\appendix
\section{$t_{\pi K^{(*)}}^{I},\, t_{\eta K^{(*)}}^{I}$ amplitudes}
\subsection{$t_{\pi K^{(*)}}^{I}$ amplitudes}
For the $\pi^0 f_1(1285)$ system, we need the $t_{\pi K^{*}}^{I=3/2}$, $t_{\pi K^{*}}^{I=1/2}$, $t_{\pi K}^{I=3/2}$ and $t_{\pi K}^{I=1/2}$ amplitudes appearing in $t_1,\, t_2$ of Eq.~\eqref{eq:4}.

First, we evaluate the amplitudes for $\pi^0 \bar K$ scattering in the following way.
Due to charge conjugation symmetry, the $\pi \bar{K}$ interaction is equivalent to the $\pi K$ interaction.

We adopt the framework developed in Ref.~\cite{Toledo:2020zxj}, which considers the interactions among the $\pi^- K^+$, $\pi^0 K^0$, and $\eta K^0$ coupled channels. Within this approach, the corresponding transition potentials $V_{ij}$ are summarized in Table~\ref{Tab:1}, where $f=93\mev$.
\begin{table*}[t]
	\centering
	\caption{Transition potentials $V_{ij}$ for the $\pi^- K^+$, $\pi^0 K^0$, and $\eta K^0$ channels.}
	\label{Tab:1}
	\setlength{\tabcolsep}{3pt} 
	\resizebox{\textwidth}{!}{ 
		\begin{tabular}{cccc}
			\hline\hline 
			\addlinespace[1ex] 
			& $\pi^- K^+$ & $\pi^0 K^0$ & $\eta K^0$ \\
			\addlinespace[1ex] 
			\midrule 
			\addlinespace[2ex] 
			
			$\pi^- K^+$ 
			& $\dfrac{-1}{6f^2} \left[ \dfrac{3}{2}s - \dfrac{3}{2s}\left(m_{\pi}^2 - m_{K}^2\right)^2 \right]$ 
			& $\begin{aligned} 
				&\dfrac{1}{2\sqrt{2}f^2} \Biggl[ \dfrac{3}{2}s - m_{\pi}^2 - m_{K}^2 \\[-1ex] 
				&\quad - \dfrac{\left(m_{\pi}^2 - m_{K}^2\right)^2}{2s} \Biggr] 
			\end{aligned}$
			& $\begin{aligned} 
				&\dfrac{1}{2\sqrt{6}f^2} \Biggl[ \dfrac{3}{2}s - \dfrac{7}{6}m_{\pi}^2 - \dfrac{1}{2}m_{\eta}^2 - \dfrac{1}{3}m_{K}^2 \\[-1ex] 
				&\quad + \dfrac{3}{2s}\left(m_{\pi}^2 - m_{K}^2\right)\left(m_{\eta}^2 - m_{K}^2\right) \Biggr] 
			\end{aligned}$ \\[7ex]
			
			$\pi^0 K^0$ 
			&  
			& $\begin{aligned} 
				&\dfrac{-1}{4f^2} \Biggl[ -\dfrac{s}{2} + m_{\pi}^2 + m_{K}^2 \\[-1ex] 
				&\quad - \dfrac{\left(m_{\pi}^2 - m_{K}^2\right)^2}{2s} \Biggr] 
			\end{aligned}$
			& $\begin{aligned} 
				&-\dfrac{1}{4\sqrt{3}f^2} \Biggl[ \dfrac{3}{2}s - \dfrac{7}{6}m_{\pi}^2 - \dfrac{1}{2}m_{\eta}^2 - \dfrac{1}{3}m_{K}^2 \\[-1ex] 
				&\quad + \dfrac{3}{2s}\left(m_{\pi}^2 - m_{K}^2\right)\left(m_{\eta}^2 - m_{K}^2\right) \Biggr] 
			\end{aligned}$ \\[7ex]
			
			$\eta K^0$ 
			&  
			&  
			& $\begin{aligned} 
				&-\dfrac{1}{4f^2} \Biggl[ -\dfrac{3}{2}s - \dfrac{2}{3}m_{\pi}^2 + m_{\eta}^2 + 3m_{K}^2 \\[-1ex] 
				&\quad - \dfrac{3}{2s}\left(m_{\eta}^2 - m_{K}^2\right)^2 \Biggr] 
			\end{aligned}$ \\[6ex]
			\hline\hline 
		\end{tabular}
	} 
\end{table*}
Subsequently, the scattering matrix is evaluated by solving the Bethe-Salpeter equation in matrix form
\begin{equation}\label{eq:A1}
	T=\left[1-VG\right]^{-1}V,
\end{equation}
where $G$ stands for the diagonal meson-meson loop function, $G=\mathrm{diag}\left[ G_i \right]$, with $G_i$ given by
\begin{equation}\label{eq:A2}
	\begin{aligned}[b]
		G_i(s) = \int_{|{\vec q\,}| < q_{\mathrm{ max}}} \,& \dfrac{\dd^3 q}{(2\pi)^3}\,\dfrac{\omega^{(i)}_1(\vec q\,) + \omega^{(i)}_2(\vec q\,)}{2 \,\omega^{(i)}_1(\vec q\, ) \, \omega^{(i)}_2(\vec q\,)} \\[2mm]
		&\times\dfrac{1}{s-[\omega^{(i)}_1(\vec q\,) + \omega^{(i)}_2(\vec q\,)]^2+i\epsilon},
	\end{aligned}
\end{equation}
where $\omega^{(i)}_j(\vec q\,) = \sqrt{{\vec{q}}^{\,2}+{m^{(i)}_{j}}^{2}}$ ($j=1,\,2$) denotes the energy of the two mesons in channel $i$, $q_{\rm max}$ is a cutoff of three-momentum, taken as $q_{\rm max}=600\mev$ from Ref.~\cite{Toledo:2020zxj}.

With the isospin phase conventions $(-\pi^+,\,\pi^0,\,\pi^-)$ and $(K^{+},\,K^{0})$, we have
\begin{align}\label{eq:A3}
	\left| \pi K ,\,I=1/2 ,\,I_3=-1/2 \right>&=\dfrac{1}{\sqrt{3}} \left| \pi^0 K^0 \right> - \sqrt{\dfrac{2}{3}} \left| \pi^- K^+ \right>,\nonumber \\[2mm]
	\left| \pi K ,\,I=3/2 ,\,I_3=-1/2 \right>&=\sqrt{\dfrac{2}{3}} \left| \pi^0 K^0 \right> + \dfrac{1}{\sqrt{3}} \left| \pi^- K^+ \right>.
\end{align}
Consequently, the corresponding isospin-projected amplitudes are given by
\begin{align}\label{eq:A4}
	t_{\pi K}^{\,I=1/2}
	&=\dfrac{1}{3}\,t_{\pi^0K^{0},\,\pi^0K^{0}}
	+\dfrac{2}{3}\,t_{\pi^-K^{+},\,\pi^-K^{+}} \nonumber \\[1mm]
	&\quad -\dfrac{2\sqrt{2}}{3}\,t_{\pi^0K^{0},\,\pi^-K^{+}},
\end{align}
\begin{align}\label{eq:A5}
	t_{\pi K}^{\,I=3/2}
	&=\dfrac{2}{3}\,t_{\pi^0K^{0},\,\pi^0K^{0}}
	+\dfrac{1}{3}\,t_{\pi^-K^{+},\,\pi^-K^{+}} \nonumber \\[1mm]
	&\quad +\dfrac{2\sqrt{2}}{3}\,t_{\pi^0K^{0},\,\pi^-K^{+}}.
\end{align}

For the $\pi K^*$ scattering, we note that the vector meson $K^*$ shares the identical isospin multiplet structure with the $K$ meson. The Clebsch-Gordan coefficients for the $\pi K^*$ coupled channels are therefore identical to those of the $\pi K$ system, allowing us to evaluate the isospin amplitudes by a direct replacement of $K$ with $K^*$ in Eqs.~\eqref{eq:A4}, \eqref{eq:A5} and in the potentials $V_{ij}$ of Table~\ref{Tab:1}, yielding
\begin{align}\label{eq:A6}
	t_{\pi K^*}^{\,I=1/2}
	&=\frac{1}{3}\,t_{\pi^0K^{*0},\,\pi^0K^{*0}}
	+\frac{2}{3}\,t_{\pi^-K^{*+},\,\pi^-K^{*+}} \nonumber \\[1mm]
	&\quad -\frac{2\sqrt{2}}{3}\,t_{\pi^0K^{*0},\,\pi^-K^{*+}},
\end{align}
\begin{align}\label{eq:A7}
	t_{\pi K^*}^{\,I=3/2}
	&=\frac{2}{3}\,t_{\pi^0K^{*0},\,\pi^0K^{*0}}
	+\frac{1}{3}\,t_{\pi^-K^{*+},\,\pi^-K^{*+}} \nonumber \\[1mm]
	&\quad +\frac{2\sqrt{2}}{3}\,t_{\pi^0K^{*0},\,\pi^-K^{*+}}.
\end{align}

The $t$ matrices for $\pi K^*$ are obtained using the potential of Table~\ref{Tab:1}, changing $K \to K^*$. 
There is an extra factor $\vec{\epsilon} \cdot \vec{\epsilon}\,'$, for the polarizations of the initial and final $K^*$, which can be ignored in the calculations, since it is $1$ for the spin allowed transitions. 
We use a convolved $G$ function for the $\pi K^*$ loop to take into account the width of the $K^*$, as done in Ref.~\cite{Jia:2025obs}.

\subsection{$t_{\eta K^{(*)}}^{I}$ amplitudes}
For the $\eta f_1(1285)$ system, the $t_{\eta K}^{I=1/2}$ and $t_{\eta K^{*}}^{I=1/2}$ amplitudes, which appear in $t_1,\, t_2$ of Eq.~\eqref{eq:3}, are needed.

To calculate the $\eta K$ scattering amplitude, we rely on the same coupled-channel unitary approach. 
The transition potentials for the $\eta K$ system are evaluated in Ref.~\cite{Toledo:2020zxj} alongside the $\pi K$ channels.
Therefore, the scattering matrix is obtained by solving the Bethe-Salpeter equation with the respective kinematics.
Since the $\eta$ meson is an isoscalar ($I=0$), the $\eta \bar{K}$ system purely couples to the $I=1/2$ sector, yielding
\begin{align}\label{eq:A8}
	t_{\eta K}^{\,I=1/2}=t_{\eta K^{0},\,\eta K^{0}}.
\end{align}

For the $\eta K^*$ system, sharing the exact isospin multiplet structure, the amplitude is obtained via the substitution $K \to K^*$,
\begin{align}\label{eq:A9}
	t_{\eta K^*}^{\,I=1/2}=t_{\eta K^{*0},\,\eta K^{*0}},
\end{align}
and the same comment as in the $\pi K^*$ $t$ matrix can be made about the $K^*$ polarization. As in the $\pi K$, $\pi K^*$ amplitudes, we use $q_{\rm max}=600\mev$. In the case of the $K^*$ amplitudes, we use the convolved $G$ function to account for the $K^*$ width.

\bibliographystyle{a}
\bibliography{refs}

\begin{thebibliography}{56}%
\makeatletter
\providecommand \@ifxundefined [1]{%
 \@ifx{#1\undefined}
}%
\providecommand \@ifnum [1]{%
 \ifnum #1\expandafter \@firstoftwo
 \else \expandafter \@secondoftwo
 \fi
}%
\providecommand \@ifx [1]{%
 \ifx #1\expandafter \@firstoftwo
 \else \expandafter \@secondoftwo
 \fi
}%
\providecommand \natexlab [1]{#1}%
\providecommand \enquote  [1]{``#1''}%
\providecommand \bibnamefont  [1]{#1}%
\providecommand \bibfnamefont [1]{#1}%
\providecommand \citenamefont [1]{#1}%
\providecommand \href@noop [0]{\@secondoftwo}%
\providecommand \href [0]{\begingroup \@sanitize@url \@href}%
\providecommand \@href[1]{\@@startlink{#1}\@@href}%
\providecommand \@@href[1]{\endgroup#1\@@endlink}%
\providecommand \@sanitize@url [0]{\catcode `\\12\catcode `\$12\catcode `\&12\catcode `\#12\catcode `\^12\catcode `\_12\catcode `\%12\relax}%
\providecommand \@@startlink[1]{}%
\providecommand \@@endlink[0]{}%
\providecommand \url  [0]{\begingroup\@sanitize@url \@url }%
\providecommand \@url [1]{\endgroup\@href {#1}{\urlprefix }}%
\providecommand \urlprefix  [0]{URL }%
\providecommand \Eprint [0]{\href }%
\providecommand \doibase [0]{https://doi.org/}%
\providecommand \selectlanguage [0]{\@gobble}%
\providecommand \bibinfo  [0]{\@secondoftwo}%
\providecommand \bibfield  [0]{\@secondoftwo}%
\providecommand \translation [1]{[#1]}%
\providecommand \BibitemOpen [0]{}%
\providecommand \bibitemStop [0]{}%
\providecommand \bibitemNoStop [0]{.\EOS\space}%
\providecommand \EOS [0]{\spacefactor3000\relax}%
\providecommand \BibitemShut  [1]{\csname bibitem#1\endcsname}%
\let\auto@bib@innerbib\@empty
\bibitem [{\citenamefont {Ablikim}\ {\it et~al.}(2022)\citenamefont {Ablikim} {\it et~al.}}]{BESIII:2022iwi}%
  \BibitemOpen
  \bibfield  {author} {\bibinfo {author} {\bibfnamefont {M.}~\bibnamefont {Ablikim}} {\it et~al.} (\bibinfo {collaboration} {BESIII}),\ }\bibinfo {title} {{Partial wave analysis of $J/\psi \to\gamma\eta\eta'$}},\ \href {https://doi.org/10.1103/PhysRevD.106.072012} {\bibfield  {journal} {\bibinfo  {journal} {Phys. Rev. D}\ }\textbf {\bibinfo {volume} {106}},\ \bibinfo {pages} {072012} (\bibinfo {year} {2022})},\ \bibinfo {note} {[Erratum: Phys.Rev.D 107, 079901 (2023)]},\ \Eprint {https://arxiv.org/abs/2202.00623} {arXiv:2202.00623 [hep-ex]} \BibitemShut {NoStop}%
\bibitem [{\citenamefont {Navas}\ {\it et~al.}(2024)\citenamefont {Navas} {\it et~al.}}]{ParticleDataGroup:2024cfk}%
  \BibitemOpen
  \bibfield  {author} {\bibinfo {author} {\bibfnamefont {S.}~\bibnamefont {Navas}} {\it et~al.} (\bibinfo {collaboration} {Particle Data Group}),\ }\bibinfo {title} {{Review of particle physics}},\ \href {https://doi.org/10.1103/PhysRevD.110.030001} {\bibfield  {journal} {\bibinfo  {journal} {Phys. Rev. D}\ }\textbf {\bibinfo {volume} {110}},\ \bibinfo {pages} {030001} (\bibinfo {year} {2024})}\BibitemShut {NoStop}%
\bibitem [{\citenamefont {Zhang}\ {\it et~al.}(2002)\citenamefont {Zhang}, \citenamefont {Ding}, \citenamefont {Li},\ and\ \citenamefont {Page}}]{Zhang:2001sb}%
  \BibitemOpen
  \bibfield  {author} {\bibinfo {author} {\bibfnamefont {R.}~\bibnamefont {Zhang}}, \bibinfo {author} {\bibfnamefont {Y.-B.}\ \bibnamefont {Ding}}, \bibinfo {author} {\bibfnamefont {X.-Q.}\ \bibnamefont {Li}},\ and\ \bibinfo {author} {\bibfnamefont {P.~R.}\ \bibnamefont {Page}},\ }\bibinfo {title} {{Molecular states and $1^{-+}$ exotic mesons}},\ \href {https://doi.org/10.1103/PhysRevD.65.096005} {\bibfield  {journal} {\bibinfo  {journal} {Phys. Rev. D}\ }\textbf {\bibinfo {volume} {65}},\ \bibinfo {pages} {096005} (\bibinfo {year} {2002})},\ \Eprint {https://arxiv.org/abs/hep-ph/0111361} {arXiv:hep-ph/0111361} \BibitemShut {NoStop}%
\bibitem [{\citenamefont {Bernard}\ {\it et~al.}(2003)\citenamefont {Bernard}, \citenamefont {Burch}, \citenamefont {Gregory}, \citenamefont {Toussaint}, \citenamefont {DeTar}, \citenamefont {Osborn}, \citenamefont {Gottlieb}, \citenamefont {Heller},\ and\ \citenamefont {Sugar}}]{Bernard:2003jd}%
  \BibitemOpen
  \bibfield  {author} {\bibinfo {author} {\bibfnamefont {C.}~\bibnamefont {Bernard}}, \bibinfo {author} {\bibfnamefont {T.}~\bibnamefont {Burch}}, \bibinfo {author} {\bibfnamefont {E.~B.}\ \bibnamefont {Gregory}}, \bibinfo {author} {\bibfnamefont {D.}~\bibnamefont {Toussaint}}, \bibinfo {author} {\bibfnamefont {C.~E.}\ \bibnamefont {DeTar}}, \bibinfo {author} {\bibfnamefont {J.}~\bibnamefont {Osborn}}, \bibinfo {author} {\bibfnamefont {S.~A.}\ \bibnamefont {Gottlieb}}, \bibinfo {author} {\bibfnamefont {U.~M.}\ \bibnamefont {Heller}},\ and\ \bibinfo {author} {\bibfnamefont {R.}~\bibnamefont {Sugar}},\ }\bibinfo {title} {{Lattice calculation of $1^{-+}$ hybrid mesons with improved Kogut-Susskind fermions}},\ \href {https://doi.org/10.1103/PhysRevD.68.074505} {\bibfield  {journal} {\bibinfo  {journal} {Phys. Rev. D}\ }\textbf {\bibinfo {volume} {68}},\ \bibinfo {pages} {074505} (\bibinfo {year} {2003})},\ \Eprint {https://arxiv.org/abs/hep-lat/0301024} {arXiv:hep-lat/0301024} \BibitemShut {NoStop}%
\bibitem [{\citenamefont {Close}\ and\ \citenamefont {Dudek}(2004)}]{Close:2003af}%
  \BibitemOpen
  \bibfield  {author} {\bibinfo {author} {\bibfnamefont {F.~E.}\ \bibnamefont {Close}}\ and\ \bibinfo {author} {\bibfnamefont {J.~J.}\ \bibnamefont {Dudek}},\ }\bibinfo {title} {{The 'Forbidden' decays of hybrid mesons to $\pi \rho$ can be large}},\ \href {https://doi.org/10.1103/PhysRevD.70.094015} {\bibfield  {journal} {\bibinfo  {journal} {Phys. Rev. D}\ }\textbf {\bibinfo {volume} {70}},\ \bibinfo {pages} {094015} (\bibinfo {year} {2004})},\ \Eprint {https://arxiv.org/abs/hep-ph/0308099} {arXiv:hep-ph/0308099} \BibitemShut {NoStop}%
\bibitem [{\citenamefont {Eshraim}\ {\it et~al.}(2020)\citenamefont {Eshraim}, \citenamefont {Fischer}, \citenamefont {Giacosa},\ and\ \citenamefont {Parganlija}}]{Eshraim:2020ucw}%
  \BibitemOpen
  \bibfield  {author} {\bibinfo {author} {\bibfnamefont {W.~I.}\ \bibnamefont {Eshraim}}, \bibinfo {author} {\bibfnamefont {C.~S.}\ \bibnamefont {Fischer}}, \bibinfo {author} {\bibfnamefont {F.}~\bibnamefont {Giacosa}},\ and\ \bibinfo {author} {\bibfnamefont {D.}~\bibnamefont {Parganlija}},\ }\bibinfo {title} {{Hybrid phenomenology in a chiral approach}},\ \href {https://doi.org/10.1140/epjp/s13360-020-00900-z} {\bibfield  {journal} {\bibinfo  {journal} {Eur. Phys. J. Plus}\ }\textbf {\bibinfo {volume} {135}},\ \bibinfo {pages} {945} (\bibinfo {year} {2020})},\ \Eprint {https://arxiv.org/abs/2001.06106} {arXiv:2001.06106 [hep-ph]} \BibitemShut {NoStop}%
\bibitem [{\citenamefont {Li}\ {\it et~al.}(2022)\citenamefont {Li}, \citenamefont {Chen}, \citenamefont {Jin},\ and\ \citenamefont {Chen}}]{Li:2021fwk}%
  \BibitemOpen
  \bibfield  {author} {\bibinfo {author} {\bibfnamefont {S.-H.}\ \bibnamefont {Li}}, \bibinfo {author} {\bibfnamefont {Z.-S.}\ \bibnamefont {Chen}}, \bibinfo {author} {\bibfnamefont {H.-Y.}\ \bibnamefont {Jin}},\ and\ \bibinfo {author} {\bibfnamefont {W.}~\bibnamefont {Chen}},\ }\bibinfo {title} {{Mass of $1^{-+}$ four-quark-hybrid mixed states}},\ \href {https://doi.org/10.1103/PhysRevD.105.054030} {\bibfield  {journal} {\bibinfo  {journal} {Phys. Rev. D}\ }\textbf {\bibinfo {volume} {105}},\ \bibinfo {pages} {054030} (\bibinfo {year} {2022})},\ \Eprint {https://arxiv.org/abs/2111.13897} {arXiv:2111.13897 [hep-ph]} \BibitemShut {NoStop}%
\bibitem [{\citenamefont {General}\ {\it et~al.}(2007)\citenamefont {General}, \citenamefont {Wang}, \citenamefont {Cotanch},\ and\ \citenamefont {Llanes-Estrada}}]{General:2007bk}%
  \BibitemOpen
  \bibfield  {author} {\bibinfo {author} {\bibfnamefont {I.~J.}\ \bibnamefont {General}}, \bibinfo {author} {\bibfnamefont {P.}~\bibnamefont {Wang}}, \bibinfo {author} {\bibfnamefont {S.~R.}\ \bibnamefont {Cotanch}},\ and\ \bibinfo {author} {\bibfnamefont {F.~J.}\ \bibnamefont {Llanes-Estrada}},\ }\bibinfo {title} {{Light $1^{-+}$ exotics: Molecular resonances}},\ \href {https://doi.org/10.1016/j.physletb.2007.08.015} {\bibfield  {journal} {\bibinfo  {journal} {Phys. Lett. B}\ }\textbf {\bibinfo {volume} {653}},\ \bibinfo {pages} {216} (\bibinfo {year} {2007})},\ \Eprint {https://arxiv.org/abs/0707.1286} {arXiv:0707.1286 [hep-ph]} \BibitemShut {NoStop}%
\bibitem [{\citenamefont {Hedditch}\ {\it et~al.}(2005)\citenamefont {Hedditch}, \citenamefont {Kamleh}, \citenamefont {Lasscock}, \citenamefont {Leinweber}, \citenamefont {Williams},\ and\ \citenamefont {Zanotti}}]{Hedditch:2005zf}%
  \BibitemOpen
  \bibfield  {author} {\bibinfo {author} {\bibfnamefont {J.~N.}\ \bibnamefont {Hedditch}}, \bibinfo {author} {\bibfnamefont {W.}~\bibnamefont {Kamleh}}, \bibinfo {author} {\bibfnamefont {B.~G.}\ \bibnamefont {Lasscock}}, \bibinfo {author} {\bibfnamefont {D.~B.}\ \bibnamefont {Leinweber}}, \bibinfo {author} {\bibfnamefont {A.~G.}\ \bibnamefont {Williams}},\ and\ \bibinfo {author} {\bibfnamefont {J.~M.}\ \bibnamefont {Zanotti}},\ }\bibinfo {title} {{$1^{-+}$ exotic meson at light quark masses}},\ \href {https://doi.org/10.1103/PhysRevD.72.114507} {\bibfield  {journal} {\bibinfo  {journal} {Phys. Rev. D}\ }\textbf {\bibinfo {volume} {72}},\ \bibinfo {pages} {114507} (\bibinfo {year} {2005})},\ \Eprint {https://arxiv.org/abs/hep-lat/0509106} {arXiv:hep-lat/0509106} \BibitemShut {NoStop}%
\bibitem [{\citenamefont {Chen}\ {\it et~al.}(2008)\citenamefont {Chen}, \citenamefont {Hosaka},\ and\ \citenamefont {Zhu}}]{Chen:2008qw}%
  \BibitemOpen
  \bibfield  {author} {\bibinfo {author} {\bibfnamefont {H.-X.}\ \bibnamefont {Chen}}, \bibinfo {author} {\bibfnamefont {A.}~\bibnamefont {Hosaka}},\ and\ \bibinfo {author} {\bibfnamefont {S.-L.}\ \bibnamefont {Zhu}},\ }\bibinfo {title} {{The $I^G\; J^{PC} = 1^-\; 1^{-+}$ Tetraquark States}},\ \href {https://doi.org/10.1103/PhysRevD.78.054017} {\bibfield  {journal} {\bibinfo  {journal} {Phys. Rev. D}\ }\textbf {\bibinfo {volume} {78}},\ \bibinfo {pages} {054017} (\bibinfo {year} {2008})},\ \Eprint {https://arxiv.org/abs/0806.1998} {arXiv:0806.1998 [hep-ph]} \BibitemShut {NoStop}%
\bibitem [{\citenamefont {Narison}(2009)}]{Narison:2009vj}%
  \BibitemOpen
  \bibfield  {author} {\bibinfo {author} {\bibfnamefont {S.}~\bibnamefont {Narison}},\ }\bibinfo {title} {{$1^{-+}$ light exotic mesons in QCD}},\ \href {https://doi.org/10.1016/j.physletb.2009.04.012} {\bibfield  {journal} {\bibinfo  {journal} {Phys. Lett. B}\ }\textbf {\bibinfo {volume} {675}},\ \bibinfo {pages} {319} (\bibinfo {year} {2009})},\ \Eprint {https://arxiv.org/abs/0903.2266} {arXiv:0903.2266 [hep-ph]} \BibitemShut {NoStop}%
\bibitem [{\citenamefont {Zhang}\ {\it et~al.}(2014)\citenamefont {Zhang}, \citenamefont {Jin},\ and\ \citenamefont {Steele}}]{Zhang:2013rya}%
  \BibitemOpen
  \bibfield  {author} {\bibinfo {author} {\bibfnamefont {Z.-f.}\ \bibnamefont {Zhang}}, \bibinfo {author} {\bibfnamefont {H.-y.}\ \bibnamefont {Jin}},\ and\ \bibinfo {author} {\bibfnamefont {T.~G.}\ \bibnamefont {Steele}},\ }\bibinfo {title} {{Revisiting $1^{-+}$ and $0^{++}$ light hybrids from Monte-Carlo based QCD sum rules}},\ \href {https://doi.org/10.1088/0256-307X/31/5/051201} {\bibfield  {journal} {\bibinfo  {journal} {Chin. Phys. Lett.}\ }\textbf {\bibinfo {volume} {31}},\ \bibinfo {pages} {051201} (\bibinfo {year} {2014})},\ \Eprint {https://arxiv.org/abs/1312.5432} {arXiv:1312.5432 [hep-ph]} \BibitemShut {NoStop}%
\bibitem [{\citenamefont {Zhang}\ {\it et~al.}(2017)\citenamefont {Zhang}, \citenamefont {Xie},\ and\ \citenamefont {Chen}}]{Zhang:2016bmy}%
  \BibitemOpen
  \bibfield  {author} {\bibinfo {author} {\bibfnamefont {X.}~\bibnamefont {Zhang}}, \bibinfo {author} {\bibfnamefont {J.-J.}\ \bibnamefont {Xie}},\ and\ \bibinfo {author} {\bibfnamefont {X.}~\bibnamefont {Chen}},\ }\bibinfo {title} {{Faddeev fixed center approximation to $\pi \bar{K} K^*$ system and the $\pi_1(1600)$}},\ \href {https://doi.org/10.1103/PhysRevD.95.056014} {\bibfield  {journal} {\bibinfo  {journal} {Phys. Rev. D}\ }\textbf {\bibinfo {volume} {95}},\ \bibinfo {pages} {056014} (\bibinfo {year} {2017})},\ \Eprint {https://arxiv.org/abs/1612.02613} {arXiv:1612.02613 [hep-ph]} \BibitemShut {NoStop}%
\bibitem [{\citenamefont {Meyer}\ and\ \citenamefont {Swanson}(2015)}]{Meyer:2015eta}%
  \BibitemOpen
  \bibfield  {author} {\bibinfo {author} {\bibfnamefont {C.~A.}\ \bibnamefont {Meyer}}\ and\ \bibinfo {author} {\bibfnamefont {E.~S.}\ \bibnamefont {Swanson}},\ }\bibinfo {title} {{Hybrid Mesons}},\ \href {https://doi.org/10.1016/j.ppnp.2015.03.001} {\bibfield  {journal} {\bibinfo  {journal} {Prog. Part. Nucl. Phys.}\ }\textbf {\bibinfo {volume} {82}},\ \bibinfo {pages} {21} (\bibinfo {year} {2015})},\ \Eprint {https://arxiv.org/abs/1502.07276} {arXiv:1502.07276 [hep-ph]} \BibitemShut {NoStop}%
\bibitem [{\citenamefont {Chen}\ {\it et~al.}(2023{\natexlab{a}})\citenamefont {Chen}, \citenamefont {Chen}, \citenamefont {Liu}, \citenamefont {Liu},\ and\ \citenamefont {Zhu}}]{Chen:2022asf}%
  \BibitemOpen
  \bibfield  {author} {\bibinfo {author} {\bibfnamefont {H.-X.}\ \bibnamefont {Chen}}, \bibinfo {author} {\bibfnamefont {W.}~\bibnamefont {Chen}}, \bibinfo {author} {\bibfnamefont {X.}~\bibnamefont {Liu}}, \bibinfo {author} {\bibfnamefont {Y.-R.}\ \bibnamefont {Liu}},\ and\ \bibinfo {author} {\bibfnamefont {S.-L.}\ \bibnamefont {Zhu}},\ }\bibinfo {title} {{An updated review of the new hadron states}},\ \href {https://doi.org/10.1088/1361-6633/aca3b6} {\bibfield  {journal} {\bibinfo  {journal} {Rept. Prog. Phys.}\ }\textbf {\bibinfo {volume} {86}},\ \bibinfo {pages} {026201} (\bibinfo {year} {2023}{\natexlab{a}})},\ \Eprint {https://arxiv.org/abs/2204.02649} {arXiv:2204.02649 [hep-ph]} \BibitemShut {NoStop}%
\bibitem [{\citenamefont {Dong}\ {\it et~al.}(2022)\citenamefont {Dong}, \citenamefont {Lin},\ and\ \citenamefont {Zou}}]{Dong:2022cuw}%
  \BibitemOpen
  \bibfield  {author} {\bibinfo {author} {\bibfnamefont {X.-K.}\ \bibnamefont {Dong}}, \bibinfo {author} {\bibfnamefont {Y.-H.}\ \bibnamefont {Lin}},\ and\ \bibinfo {author} {\bibfnamefont {B.-S.}\ \bibnamefont {Zou}},\ }\bibinfo {title} {{Interpretation of the $\eta_{1}(1855)$ as a $K\bar K_1(1400) + c.c.$ molecule}},\ \href {https://doi.org/10.1007/s11433-022-1887-5} {\bibfield  {journal} {\bibinfo  {journal} {Sci. China Phys. Mech. Astron.}\ }\textbf {\bibinfo {volume} {65}},\ \bibinfo {pages} {261011} (\bibinfo {year} {2022})},\ \Eprint {https://arxiv.org/abs/2202.00863} {arXiv:2202.00863 [hep-ph]} \BibitemShut {NoStop}%
\bibitem [{\citenamefont {Yang}\ {\it et~al.}(2023)\citenamefont {Yang}, \citenamefont {Zhu},\ and\ \citenamefont {Huang}}]{Yang:2022rck}%
  \BibitemOpen
  \bibfield  {author} {\bibinfo {author} {\bibfnamefont {F.}~\bibnamefont {Yang}}, \bibinfo {author} {\bibfnamefont {H.~Q.}\ \bibnamefont {Zhu}},\ and\ \bibinfo {author} {\bibfnamefont {Y.}~\bibnamefont {Huang}},\ }\bibinfo {title} {{Analysis of the $\eta_1(1855)$ as a $K \bar K_1(1400)$ molecular state}},\ \href {https://doi.org/10.1016/j.nuclphysa.2022.122571} {\bibfield  {journal} {\bibinfo  {journal} {Nucl. Phys. A}\ }\textbf {\bibinfo {volume} {1030}},\ \bibinfo {pages} {122571} (\bibinfo {year} {2023})},\ \Eprint {https://arxiv.org/abs/2203.06934} {arXiv:2203.06934 [hep-ph]} \BibitemShut {NoStop}%
\bibitem [{\citenamefont {Huang}\ and\ \citenamefont {Zhu}(2023)}]{Huang:2022tpq}%
  \BibitemOpen
  \bibfield  {author} {\bibinfo {author} {\bibfnamefont {Y.}~\bibnamefont {Huang}}\ and\ \bibinfo {author} {\bibfnamefont {H.~Q.}\ \bibnamefont {Zhu}},\ }\bibinfo {title} {{Revealing the inner structure of the newly observed $\eta_1(1855)$ via photoproduction}},\ \href {https://doi.org/10.1088/1361-6471/ace4e2} {\bibfield  {journal} {\bibinfo  {journal} {J. Phys. G}\ }\textbf {\bibinfo {volume} {50}},\ \bibinfo {pages} {095002} (\bibinfo {year} {2023})},\ \Eprint {https://arxiv.org/abs/2209.02879} {arXiv:2209.02879 [hep-ph]} \BibitemShut {NoStop}%
\bibitem [{\citenamefont {Liu}\ {\it et~al.}(2025)\citenamefont {Liu}, \citenamefont {Chen}, \citenamefont {Lian}, \citenamefont {Li},\ and\ \citenamefont {Chen}}]{Liu:2024lph}%
  \BibitemOpen
  \bibfield  {author} {\bibinfo {author} {\bibfnamefont {Z.-S.}\ \bibnamefont {Liu}}, \bibinfo {author} {\bibfnamefont {X.-L.}\ \bibnamefont {Chen}}, \bibinfo {author} {\bibfnamefont {D.-K.}\ \bibnamefont {Lian}}, \bibinfo {author} {\bibfnamefont {N.}~\bibnamefont {Li}},\ and\ \bibinfo {author} {\bibfnamefont {W.}~\bibnamefont {Chen}},\ }\bibinfo {title} {{Mixing angle of $K_1(1270/1400)$ and the $K \bar K_1(1400)$ molecular interpretation of $\eta_1(1855)$}},\ \href {https://doi.org/10.1103/PhysRevD.111.014014} {\bibfield  {journal} {\bibinfo  {journal} {Phys. Rev. D}\ }\textbf {\bibinfo {volume} {111}},\ \bibinfo {pages} {014014} (\bibinfo {year} {2025})},\ \Eprint {https://arxiv.org/abs/2411.01867} {arXiv:2411.01867 [hep-ph]} \BibitemShut {NoStop}%
\bibitem [{\citenamefont {Shastry}\ {\it et~al.}(2022)\citenamefont {Shastry}, \citenamefont {Fischer},\ and\ \citenamefont {Giacosa}}]{Shastry:2022mhk}%
  \BibitemOpen
  \bibfield  {author} {\bibinfo {author} {\bibfnamefont {V.}~\bibnamefont {Shastry}}, \bibinfo {author} {\bibfnamefont {C.~S.}\ \bibnamefont {Fischer}},\ and\ \bibinfo {author} {\bibfnamefont {F.}~\bibnamefont {Giacosa}},\ }\bibinfo {title} {{The phenomenology of the exotic hybrid nonet with $\pi_1(1600)$ and $\eta_1(1855)$}},\ \href {https://doi.org/10.1016/j.physletb.2022.137478} {\bibfield  {journal} {\bibinfo  {journal} {Phys. Lett. B}\ }\textbf {\bibinfo {volume} {834}},\ \bibinfo {pages} {137478} (\bibinfo {year} {2022})},\ \Eprint {https://arxiv.org/abs/2203.04327} {arXiv:2203.04327 [hep-ph]} \BibitemShut {NoStop}%
\bibitem [{\citenamefont {Chen}\ {\it et~al.}(2023{\natexlab{b}})\citenamefont {Chen}, \citenamefont {Luo},\ and\ \citenamefont {Liu}}]{Chen:2023ukh}%
  \BibitemOpen
  \bibfield  {author} {\bibinfo {author} {\bibfnamefont {B.}~\bibnamefont {Chen}}, \bibinfo {author} {\bibfnamefont {S.-Q.}\ \bibnamefont {Luo}},\ and\ \bibinfo {author} {\bibfnamefont {X.}~\bibnamefont {Liu}},\ }\bibinfo {title} {{Constructing the $J^{P(C)}=1^{-(+)}$ light flavor hybrid nonet with the newly observed $\eta_1(1855)$}},\ \href {https://doi.org/10.1103/PhysRevD.108.054034} {\bibfield  {journal} {\bibinfo  {journal} {Phys. Rev. D}\ }\textbf {\bibinfo {volume} {108}},\ \bibinfo {pages} {054034} (\bibinfo {year} {2023}{\natexlab{b}})},\ \Eprint {https://arxiv.org/abs/2302.06785} {arXiv:2302.06785 [hep-ph]} \BibitemShut {NoStop}%
\bibitem [{\citenamefont {Tan}\ {\it et~al.}(2024)\citenamefont {Tan}, \citenamefont {Su},\ and\ \citenamefont {Chen}}]{Tan:2024grd}%
  \BibitemOpen
  \bibfield  {author} {\bibinfo {author} {\bibfnamefont {W.-H.}\ \bibnamefont {Tan}}, \bibinfo {author} {\bibfnamefont {N.}~\bibnamefont {Su}},\ and\ \bibinfo {author} {\bibfnamefont {H.-X.}\ \bibnamefont {Chen}},\ }\bibinfo {title} {{Light single-gluon hybrid states with various exotic quantum numbers}},\ \href {https://doi.org/10.1103/PhysRevD.110.034031} {\bibfield  {journal} {\bibinfo  {journal} {Phys. Rev. D}\ }\textbf {\bibinfo {volume} {110}},\ \bibinfo {pages} {034031} (\bibinfo {year} {2024})},\ \Eprint {https://arxiv.org/abs/2404.09538} {arXiv:2404.09538 [hep-ph]} \BibitemShut {NoStop}%
\bibitem [{\citenamefont {Qiu}\ and\ \citenamefont {Zhao}(2022)}]{Qiu:2022ktc}%
  \BibitemOpen
  \bibfield  {author} {\bibinfo {author} {\bibfnamefont {L.}~\bibnamefont {Qiu}}\ and\ \bibinfo {author} {\bibfnamefont {Q.}~\bibnamefont {Zhao}},\ }\bibinfo {title} {{Towards the establishment of the light $J^{P (C )}=1^{-(+)}$ hybrid nonet}},\ \href {https://doi.org/10.1088/1674-1137/ac567e} {\bibfield  {journal} {\bibinfo  {journal} {Chin. Phys. C}\ }\textbf {\bibinfo {volume} {46}},\ \bibinfo {pages} {051001} (\bibinfo {year} {2022})},\ \Eprint {https://arxiv.org/abs/2202.00904} {arXiv:2202.00904 [hep-ph]} \BibitemShut {NoStop}%
\bibitem [{\citenamefont {Zhang}\ {\it et~al.}(2026)\citenamefont {Zhang}, \citenamefont {Huang},\ and\ \citenamefont {Wang}}]{Zhang:2025xee}%
  \BibitemOpen
  \bibfield  {author} {\bibinfo {author} {\bibfnamefont {F.-Y.}\ \bibnamefont {Zhang}}, \bibinfo {author} {\bibfnamefont {Q.}~\bibnamefont {Huang}},\ and\ \bibinfo {author} {\bibfnamefont {L.-M.}\ \bibnamefont {Wang}},\ }\bibinfo {title} {{Spectral analysis and decay mechanisms of $1^{-+}$ hybrid states in light meson sector}},\ \href {https://doi.org/10.1103/1wkx-3s2l} {\bibfield  {journal} {\bibinfo  {journal} {Phys. Rev. D}\ }\textbf {\bibinfo {volume} {113}},\ \bibinfo {pages} {014002} (\bibinfo {year} {2026})},\ \Eprint {https://arxiv.org/abs/2503.01443} {arXiv:2503.01443 [hep-ph]} \BibitemShut {NoStop}%
\bibitem [{\citenamefont {Wan}\ {\it et~al.}(2022)\citenamefont {Wan}, \citenamefont {Zhang},\ and\ \citenamefont {Qiao}}]{Wan:2022xkx}%
  \BibitemOpen
  \bibfield  {author} {\bibinfo {author} {\bibfnamefont {B.-D.}\ \bibnamefont {Wan}}, \bibinfo {author} {\bibfnamefont {S.-Q.}\ \bibnamefont {Zhang}},\ and\ \bibinfo {author} {\bibfnamefont {C.-F.}\ \bibnamefont {Qiao}},\ }\bibinfo {title} {{Possible structure of the newly found exotic state $\eta_1(1855)$}},\ \href {https://doi.org/10.1103/PhysRevD.106.074003} {\bibfield  {journal} {\bibinfo  {journal} {Phys. Rev. D}\ }\textbf {\bibinfo {volume} {106}},\ \bibinfo {pages} {074003} (\bibinfo {year} {2022})},\ \Eprint {https://arxiv.org/abs/2203.14014} {arXiv:2203.14014 [hep-ph]} \BibitemShut {NoStop}%
\bibitem [{\citenamefont {Yan}\ {\it et~al.}(2023)\citenamefont {Yan}, \citenamefont {Dias}, \citenamefont {Guevara}, \citenamefont {Guo},\ and\ \citenamefont {Zou}}]{Yan:2023vbh}%
  \BibitemOpen
  \bibfield  {author} {\bibinfo {author} {\bibfnamefont {M.-J.}\ \bibnamefont {Yan}}, \bibinfo {author} {\bibfnamefont {J.~M.}\ \bibnamefont {Dias}}, \bibinfo {author} {\bibfnamefont {A.}~\bibnamefont {Guevara}}, \bibinfo {author} {\bibfnamefont {F.-K.}\ \bibnamefont {Guo}},\ and\ \bibinfo {author} {\bibfnamefont {B.-S.}\ \bibnamefont {Zou}},\ }\bibinfo {title} {{On the $\eta_1(1855)$, $\pi_1(1400)$ and $\pi_1(1600)$ as Dynamically Generated States and Their SU(3) Partners}},\ \href {https://doi.org/10.3390/universe9020109} {\bibfield  {journal} {\bibinfo  {journal} {Universe}\ }\textbf {\bibinfo {volume} {9}},\ \bibinfo {pages} {109} (\bibinfo {year} {2023})},\ \Eprint {https://arxiv.org/abs/2301.04432} {arXiv:2301.04432 [hep-ph]} \BibitemShut {NoStop}%
\bibitem [{\citenamefont {Lutz}\ and\ \citenamefont {Kolomeitsev}(2004)}]{Lutz:2003fm}%
  \BibitemOpen
  \bibfield  {author} {\bibinfo {author} {\bibfnamefont {M.~F.~M.}\ \bibnamefont {Lutz}}\ and\ \bibinfo {author} {\bibfnamefont {E.~E.}\ \bibnamefont {Kolomeitsev}},\ }\bibinfo {title} {{On meson resonances and chiral symmetry}},\ \href {https://doi.org/10.1016/j.nuclphysa.2003.11.009} {\bibfield  {journal} {\bibinfo  {journal} {Nucl. Phys. A}\ }\textbf {\bibinfo {volume} {730}},\ \bibinfo {pages} {392} (\bibinfo {year} {2004})},\ \Eprint {https://arxiv.org/abs/nucl-th/0307039} {arXiv:nucl-th/0307039} \BibitemShut {NoStop}%
\bibitem [{\citenamefont {Roca}\ {\it et~al.}(2005)\citenamefont {Roca}, \citenamefont {Oset},\ and\ \citenamefont {Singh}}]{Roca:2005nm}%
  \BibitemOpen
  \bibfield  {author} {\bibinfo {author} {\bibfnamefont {L.}~\bibnamefont {Roca}}, \bibinfo {author} {\bibfnamefont {E.}~\bibnamefont {Oset}},\ and\ \bibinfo {author} {\bibfnamefont {J.}~\bibnamefont {Singh}},\ }\bibinfo {title} {{Low lying axial-vector mesons as dynamically generated resonances}},\ \href {https://doi.org/10.1103/PhysRevD.72.014002} {\bibfield  {journal} {\bibinfo  {journal} {Phys. Rev. D}\ }\textbf {\bibinfo {volume} {72}},\ \bibinfo {pages} {014002} (\bibinfo {year} {2005})},\ \Eprint {https://arxiv.org/abs/hep-ph/0503273} {arXiv:hep-ph/0503273} \BibitemShut {NoStop}%
\bibitem [{\citenamefont {Zhou}\ {\it et~al.}(2014)\citenamefont {Zhou}, \citenamefont {Ren}, \citenamefont {Chen},\ and\ \citenamefont {Geng}}]{Zhou:2014ila}%
  \BibitemOpen
  \bibfield  {author} {\bibinfo {author} {\bibfnamefont {Y.}~\bibnamefont {Zhou}}, \bibinfo {author} {\bibfnamefont {X.-L.}\ \bibnamefont {Ren}}, \bibinfo {author} {\bibfnamefont {H.-X.}\ \bibnamefont {Chen}},\ and\ \bibinfo {author} {\bibfnamefont {L.-S.}\ \bibnamefont {Geng}},\ }\bibinfo {title} {{Pseudoscalar meson and vector meson interactions and dynamically generated axial-vector mesons}},\ \href {https://doi.org/10.1103/PhysRevD.90.014020} {\bibfield  {journal} {\bibinfo  {journal} {Phys. Rev. D}\ }\textbf {\bibinfo {volume} {90}},\ \bibinfo {pages} {014020} (\bibinfo {year} {2014})},\ \Eprint {https://arxiv.org/abs/1404.6847} {arXiv:1404.6847 [nucl-th]} \BibitemShut {NoStop}%
\bibitem [{\citenamefont {Ericson}\ and\ \citenamefont {Weise}(1988)}]{Ericson:1988gk}%
  \BibitemOpen
  \bibfield  {author} {\bibinfo {author} {\bibfnamefont {T.~E.~O.}\ \bibnamefont {Ericson}}\ and\ \bibinfo {author} {\bibfnamefont {W.}~\bibnamefont {Weise}},\ }\href@noop {} {\bibinfo {title} {{Pions and Nuclei}}}\ (\bibinfo  {publisher} {Clarendon Press},\ \bibinfo {address} {Oxford, UK},\ \bibinfo {year} {1988})\BibitemShut {NoStop}%
\bibitem [{\citenamefont {Seki}\ and\ \citenamefont {Masutani}(1983)}]{Seki:1983sh}%
  \BibitemOpen
  \bibfield  {author} {\bibinfo {author} {\bibfnamefont {R.}~\bibnamefont {Seki}}\ and\ \bibinfo {author} {\bibfnamefont {K.}~\bibnamefont {Masutani}},\ }\bibinfo {title} {{Unified analysis of pionic atoms and low-energy pion-nucleus scattering: Phenomenological analysis}},\ \href {https://doi.org/10.1103/PhysRevC.27.2799} {\bibfield  {journal} {\bibinfo  {journal} {Phys. Rev. C}\ }\textbf {\bibinfo {volume} {27}},\ \bibinfo {pages} {2799} (\bibinfo {year} {1983})}\BibitemShut {NoStop}%
\bibitem [{\citenamefont {Nieves}\ {\it et~al.}(1993)\citenamefont {Nieves}, \citenamefont {Oset},\ and\ \citenamefont {Garcia-Recio}}]{Nieves:1993ev}%
  \BibitemOpen
  \bibfield  {author} {\bibinfo {author} {\bibfnamefont {J.}~\bibnamefont {Nieves}}, \bibinfo {author} {\bibfnamefont {E.}~\bibnamefont {Oset}},\ and\ \bibinfo {author} {\bibfnamefont {C.}~\bibnamefont {Garcia-Recio}},\ }\bibinfo {title} {{A Theoretical approach to pionic atoms and the problem of anomalies}},\ \href {https://doi.org/10.1016/0375-9474(93)90245-S} {\bibfield  {journal} {\bibinfo  {journal} {Nucl. Phys. A}\ }\textbf {\bibinfo {volume} {554}},\ \bibinfo {pages} {509} (\bibinfo {year} {1993})}\BibitemShut {NoStop}%
\bibitem [{\citenamefont {Brown}\ and\ \citenamefont {Weise}(1975)}]{Brown:1975di}%
  \BibitemOpen
  \bibfield  {author} {\bibinfo {author} {\bibfnamefont {G.~E.}\ \bibnamefont {Brown}}\ and\ \bibinfo {author} {\bibfnamefont {W.}~\bibnamefont {Weise}},\ }\bibinfo {title} {{Pion Scattering and Isobars in Nuclei}},\ \href {https://doi.org/10.1016/0370-1573(75)90026-5} {\bibfield  {journal} {\bibinfo  {journal} {Phys. Rept.}\ }\textbf {\bibinfo {volume} {22}},\ \bibinfo {pages} {279} (\bibinfo {year} {1975})}\BibitemShut {NoStop}%
\bibitem [{\citenamefont {\v{S}erk\v{s}nyt\.{e}}\ and\ \citenamefont {Kundu}()}]{Lauraser}%
  \BibitemOpen
  \bibfield  {author} {\bibinfo {author} {\bibfnamefont {L.}~\bibnamefont {\v{S}erk\v{s}nyt\.{e}}}\ and\ \bibinfo {author} {\bibfnamefont {S.}~\bibnamefont {Kundu}},\ }\href@noop {} {\bibinfo {title} {First experimental study of axial-vector meson-nucleon interactions using $p$-$f_1 (1285)$ correlations with {ALICE}}},\ \bibinfo {howpublished} {\href{https://indico.global/event/13943/contributions/143540/attachments/68255/132344/2026-03-24-SQM_f1p_v5.pdf}{(talk given by Laura \v{S}erk\v{s}nyt\.{e} in the Strangeness in Quark Matter Conference, Los Angeles, March 2026)}}\BibitemShut {NoStop}%
\bibitem [{\citenamefont {Encarnaci{\'o}n}\ {\it et~al.}(2025)\citenamefont {Encarnaci{\'o}n}, \citenamefont {Feijoo},\ and\ \citenamefont {Oset}}]{Encarnacion:2025lyf}%
  \BibitemOpen
  \bibfield  {author} {\bibinfo {author} {\bibfnamefont {P.}~\bibnamefont {Encarnaci{\'o}n}}, \bibinfo {author} {\bibfnamefont {A.}~\bibnamefont {Feijoo}},\ and\ \bibinfo {author} {\bibfnamefont {E.}~\bibnamefont {Oset}},\ }\bibinfo {title} {{Correlation function for the $p f_1(1285)$ interaction}},\ \href {https://doi.org/10.1103/s7g4-6lmv} {\bibfield  {journal} {\bibinfo  {journal} {Phys. Rev. D}\ }\textbf {\bibinfo {volume} {111}},\ \bibinfo {pages} {114023} (\bibinfo {year} {2025})},\ \Eprint {https://arxiv.org/abs/2502.19329} {arXiv:2502.19329 [hep-ph]} \BibitemShut {NoStop}%
\bibitem [{\citenamefont {Ikeno}\ and\ \citenamefont {Oset}(2025)}]{Ikeno:2025bsx}%
  \BibitemOpen
  \bibfield  {author} {\bibinfo {author} {\bibfnamefont {N.}~\bibnamefont {Ikeno}}\ and\ \bibinfo {author} {\bibfnamefont {E.}~\bibnamefont {Oset}},\ }\bibinfo {title} {{Correlation function for the $n \bar D_{s0}^*(2317)$ interaction and the issue of elastic unitarity}},\ \href {https://doi.org/10.1103/bb31-rdjb} {\bibfield  {journal} {\bibinfo  {journal} {Phys. Rev. D}\ }\textbf {\bibinfo {volume} {112}},\ \bibinfo {pages} {094019} (\bibinfo {year} {2025})},\ \Eprint {https://arxiv.org/abs/2507.16367} {arXiv:2507.16367 [hep-ph]} \BibitemShut {NoStop}%
\bibitem [{\citenamefont {Agat{\~a}o}\ {\it et~al.}(2025)\citenamefont {Agat{\~a}o}, \citenamefont {Brand{\~a}o}, \citenamefont {Mart{\'\i}nez~Torres}, \citenamefont {Khemchandani}, \citenamefont {Abreu},\ and\ \citenamefont {Oset}}]{Agatao:2025ckp}%
  \BibitemOpen
  \bibfield  {author} {\bibinfo {author} {\bibfnamefont {B.}~\bibnamefont {Agat{\~a}o}}, \bibinfo {author} {\bibfnamefont {P.}~\bibnamefont {Brand{\~a}o}}, \bibinfo {author} {\bibfnamefont {A.}~\bibnamefont {Mart{\'\i}nez~Torres}}, \bibinfo {author} {\bibfnamefont {K.~P.}\ \bibnamefont {Khemchandani}}, \bibinfo {author} {\bibfnamefont {L.~M.}\ \bibnamefont {Abreu}},\ and\ \bibinfo {author} {\bibfnamefont {E.}~\bibnamefont {Oset}},\ }\bibinfo {title} {{Correlation functions for $n\,\bar{D}_{s1}(2460)$ and $n\,\bar{D}_{s1}(2536)$}},\ \href {https://doi.org/10.1140/epjc/s10052-025-14838-y} {\bibfield  {journal} {\bibinfo  {journal} {Eur. Phys. J. C}\ }\textbf {\bibinfo {volume} {85}},\ \bibinfo {pages} {1136} (\bibinfo {year} {2025})},\ \Eprint {https://arxiv.org/abs/2508.05825} {arXiv:2508.05825 [hep-ph]} \BibitemShut {NoStop}%
\bibitem [{\citenamefont {Jia}\ {\it et~al.}(2026{\natexlab{a}})\citenamefont {Jia}, \citenamefont {Song}, \citenamefont {Liang},\ and\ \citenamefont {Oset}}]{Jia:2026dpl}%
  \BibitemOpen
  \bibfield  {author} {\bibinfo {author} {\bibfnamefont {W.-H.}\ \bibnamefont {Jia}}, \bibinfo {author} {\bibfnamefont {J.}~\bibnamefont {Song}}, \bibinfo {author} {\bibfnamefont {W.-H.}\ \bibnamefont {Liang}},\ and\ \bibinfo {author} {\bibfnamefont {E.}~\bibnamefont {Oset}},\ }\href@noop {} {\bibinfo {title} {{Scattering data and correlation function for the $K f_1(1285)$ interaction}}} (\bibinfo {year} {2026}{\natexlab{a}}),\ \Eprint {https://arxiv.org/abs/2602.16683} {arXiv:2602.16683 [Eur. Phys. J. C in print]} \BibitemShut {NoStop}%
\bibitem [{\citenamefont {Jia}\ {\it et~al.}(2026{\natexlab{b}})\citenamefont {Jia}, \citenamefont {Li}, \citenamefont {Liang}, \citenamefont {Song},\ and\ \citenamefont {Oset}}]{Jia:2026iqo}%
  \BibitemOpen
  \bibfield  {author} {\bibinfo {author} {\bibfnamefont {W.-H.}\ \bibnamefont {Jia}}, \bibinfo {author} {\bibfnamefont {H.-P.}\ \bibnamefont {Li}}, \bibinfo {author} {\bibfnamefont {W.-H.}\ \bibnamefont {Liang}}, \bibinfo {author} {\bibfnamefont {J.}~\bibnamefont {Song}},\ and\ \bibinfo {author} {\bibfnamefont {E.}~\bibnamefont {Oset}},\ }\href@noop {} {\bibinfo {title} {{Correlation function and bound state from the $K D_{s0}^*(2317)$ interaction}}} (\bibinfo {year} {2026}{\natexlab{b}}),\ \Eprint {https://arxiv.org/abs/2604.07261} {arXiv:2604.07261 [hep-ph]} \BibitemShut {NoStop}%
\bibitem [{\citenamefont {Encarnaci{\'o}n}\ {\it et~al.}(2026)\citenamefont {Encarnaci{\'o}n}, \citenamefont {Feijoo},\ and\ \citenamefont {Oset}}]{Encarnacion:2026zas}%
  \BibitemOpen
  \bibfield  {author} {\bibinfo {author} {\bibfnamefont {P.}~\bibnamefont {Encarnaci{\'o}n}}, \bibinfo {author} {\bibfnamefont {A.}~\bibnamefont {Feijoo}},\ and\ \bibinfo {author} {\bibfnamefont {E.}~\bibnamefont {Oset}},\ }\bibinfo {title} {{Scattering observables and correlation function for $p$ $f_1(1285)$ revisited}},\ \href {https://doi.org/10.1103/kk4c-xv4d} {\bibfield  {journal} {\bibinfo  {journal} {Phys. Rev. D}\ }\textbf {\bibinfo {volume} {113}},\ \bibinfo {pages} {L111502} (\bibinfo {year} {2026})},\ \Eprint {https://arxiv.org/abs/2603.09852} {arXiv:2603.09852 [hep-ph]} \BibitemShut {NoStop}%
\bibitem [{\citenamefont {Molina}\ {\it et~al.}(2008)\citenamefont {Molina}, \citenamefont {Nicmorus},\ and\ \citenamefont {Oset}}]{Molina:2008jw}%
  \BibitemOpen
  \bibfield  {author} {\bibinfo {author} {\bibfnamefont {R.}~\bibnamefont {Molina}}, \bibinfo {author} {\bibfnamefont {D.}~\bibnamefont {Nicmorus}},\ and\ \bibinfo {author} {\bibfnamefont {E.}~\bibnamefont {Oset}},\ }\bibinfo {title} {{The $\rho$ $\rho$ interaction in the hidden gauge formalism and the $f_0(1370)$ and $f_2(1270)$ resonances}},\ \href {https://doi.org/10.1103/PhysRevD.78.114018} {\bibfield  {journal} {\bibinfo  {journal} {Phys. Rev. D}\ }\textbf {\bibinfo {volume} {78}},\ \bibinfo {pages} {114018} (\bibinfo {year} {2008})},\ \Eprint {https://arxiv.org/abs/0809.2233} {arXiv:0809.2233 [hep-ph]} \BibitemShut {NoStop}%
\bibitem [{\citenamefont {Geng}\ and\ \citenamefont {Oset}(2009)}]{Geng:2008gx}%
  \BibitemOpen
  \bibfield  {author} {\bibinfo {author} {\bibfnamefont {L.~S.}\ \bibnamefont {Geng}}\ and\ \bibinfo {author} {\bibfnamefont {E.}~\bibnamefont {Oset}},\ }\bibinfo {title} {{Vector meson-vector meson interaction in a hidden gauge unitary approach}},\ \href {https://doi.org/10.1103/PhysRevD.79.074009} {\bibfield  {journal} {\bibinfo  {journal} {Phys. Rev. D}\ }\textbf {\bibinfo {volume} {79}},\ \bibinfo {pages} {074009} (\bibinfo {year} {2009})},\ \Eprint {https://arxiv.org/abs/0812.1199} {arXiv:0812.1199 [hep-ph]} \BibitemShut {NoStop}%
\bibitem [{\citenamefont {Aceti}\ {\it et~al.}(2015{\natexlab{a}})\citenamefont {Aceti}, \citenamefont {Dias},\ and\ \citenamefont {Oset}}]{Aceti:2015zva}%
  \BibitemOpen
  \bibfield  {author} {\bibinfo {author} {\bibfnamefont {F.}~\bibnamefont {Aceti}}, \bibinfo {author} {\bibfnamefont {J.~M.}\ \bibnamefont {Dias}},\ and\ \bibinfo {author} {\bibfnamefont {E.}~\bibnamefont {Oset}},\ }\bibinfo {title} {{$f_1(1285)$ decays into $a_0(980) \pi^0$, $f_0(980) \pi^0$ and isospin breaking}},\ \href {https://doi.org/10.1140/epja/i2015-15048-5} {\bibfield  {journal} {\bibinfo  {journal} {Eur. Phys. J. A}\ }\textbf {\bibinfo {volume} {51}},\ \bibinfo {pages} {48} (\bibinfo {year} {2015}{\natexlab{a}})},\ \Eprint {https://arxiv.org/abs/1501.06505} {arXiv:1501.06505 [hep-ph]} \BibitemShut {NoStop}%
\bibitem [{\citenamefont {Aceti}\ {\it et~al.}(2015{\natexlab{b}})\citenamefont {Aceti}, \citenamefont {Xie},\ and\ \citenamefont {Oset}}]{Aceti:2015pma}%
  \BibitemOpen
  \bibfield  {author} {\bibinfo {author} {\bibfnamefont {F.}~\bibnamefont {Aceti}}, \bibinfo {author} {\bibfnamefont {J.-J.}\ \bibnamefont {Xie}},\ and\ \bibinfo {author} {\bibfnamefont {E.}~\bibnamefont {Oset}},\ }\bibinfo {title} {{The $K \bar K \pi$ decay of the $f_1(1285)$ and its nature as a $K^* \bar K -cc$ molecule}},\ \href {https://doi.org/10.1016/j.physletb.2015.09.068} {\bibfield  {journal} {\bibinfo  {journal} {Phys. Lett. B}\ }\textbf {\bibinfo {volume} {750}},\ \bibinfo {pages} {609} (\bibinfo {year} {2015}{\natexlab{b}})},\ \Eprint {https://arxiv.org/abs/1505.06134} {arXiv:1505.06134 [hep-ph]} \BibitemShut {NoStop}%
\bibitem [{\citenamefont {Xie}(2015)}]{Xie:2015wja}%
  \BibitemOpen
  \bibfield  {author} {\bibinfo {author} {\bibfnamefont {J.-J.}\ \bibnamefont {Xie}},\ }\bibinfo {title} {{The $K^- p \to f_1(1285) \Lambda$ reaction within an effective Lagrangian approach}},\ \href {https://doi.org/10.1103/PhysRevC.92.065203} {\bibfield  {journal} {\bibinfo  {journal} {Phys. Rev. C}\ }\textbf {\bibinfo {volume} {92}},\ \bibinfo {pages} {065203} (\bibinfo {year} {2015})},\ \Eprint {https://arxiv.org/abs/1509.06196} {arXiv:1509.06196 [nucl-th]} \BibitemShut {NoStop}%
\bibitem [{\citenamefont {Xie}\ and\ \citenamefont {Oset}(2016)}]{Xie:2015lta}%
  \BibitemOpen
  \bibfield  {author} {\bibinfo {author} {\bibfnamefont {J.-J.}\ \bibnamefont {Xie}}\ and\ \bibinfo {author} {\bibfnamefont {E.}~\bibnamefont {Oset}},\ }\bibinfo {title} {{Role of the $f_1(1285)$ state in the $J/\psi \to \phi \bar{K} K^*$ and $J/\psi \to \phi f_1(1285)$ decays}},\ \href {https://doi.org/10.1016/j.physletb.2015.12.064} {\bibfield  {journal} {\bibinfo  {journal} {Phys. Lett. B}\ }\textbf {\bibinfo {volume} {753}},\ \bibinfo {pages} {591} (\bibinfo {year} {2016})},\ \Eprint {https://arxiv.org/abs/1509.08099} {arXiv:1509.08099 [hep-ph]} \BibitemShut {NoStop}%
\bibitem [{\citenamefont {Molina}\ {\it et~al.}(2016)\citenamefont {Molina}, \citenamefont {D{\"o}ring},\ and\ \citenamefont {Oset}}]{Molina:2016pbg}%
  \BibitemOpen
  \bibfield  {author} {\bibinfo {author} {\bibfnamefont {R.}~\bibnamefont {Molina}}, \bibinfo {author} {\bibfnamefont {M.}~\bibnamefont {D{\"o}ring}},\ and\ \bibinfo {author} {\bibfnamefont {E.}~\bibnamefont {Oset}},\ }\bibinfo {title} {{Determination of the compositeness of resonances from decays: the case of the $B^0_s\to J/\psi f_1(1285)$}},\ \href {https://doi.org/10.1103/PhysRevD.93.114004} {\bibfield  {journal} {\bibinfo  {journal} {Phys. Rev. D}\ }\textbf {\bibinfo {volume} {93}},\ \bibinfo {pages} {114004} (\bibinfo {year} {2016})},\ \Eprint {https://arxiv.org/abs/1604.02574} {arXiv:1604.02574 [hep-ph]} \BibitemShut {NoStop}%
\bibitem [{\citenamefont {Oset}\ and\ \citenamefont {Roca}(2018)}]{Oset:2018zgc}%
  \BibitemOpen
  \bibfield  {author} {\bibinfo {author} {\bibfnamefont {E.}~\bibnamefont {Oset}}\ and\ \bibinfo {author} {\bibfnamefont {L.}~\bibnamefont {Roca}},\ }\bibinfo {title} {{Triangle singularity in $\tau \to f_1(1285)\pi\nu_\tau$ decay}},\ \href {https://doi.org/10.1016/j.physletb.2018.05.056} {\bibfield  {journal} {\bibinfo  {journal} {Phys. Lett. B}\ }\textbf {\bibinfo {volume} {782}},\ \bibinfo {pages} {332} (\bibinfo {year} {2018})},\ \Eprint {https://arxiv.org/abs/1803.07807} {arXiv:1803.07807 [hep-ph]} \BibitemShut {NoStop}%
\bibitem [{\citenamefont {He}\ {\it et~al.}(2021)\citenamefont {He}, \citenamefont {Luo}, \citenamefont {Xie},\ and\ \citenamefont {Sun}}]{He:2021exv}%
  \BibitemOpen
  \bibfield  {author} {\bibinfo {author} {\bibfnamefont {D.}~\bibnamefont {He}}, \bibinfo {author} {\bibfnamefont {X.}~\bibnamefont {Luo}}, \bibinfo {author} {\bibfnamefont {Y.}~\bibnamefont {Xie}},\ and\ \bibinfo {author} {\bibfnamefont {H.}~\bibnamefont {Sun}},\ }\bibinfo {title} {{Theoretical study in the $\bar{B}^0 \to J/\psi \bar{K}^{*0} K^0$ and $\bar{B}^0 \to J/\psi f_1(1285)$ decays}},\ \href {https://doi.org/10.1103/PhysRevD.103.094007} {\bibfield  {journal} {\bibinfo  {journal} {Phys. Rev. D}\ }\textbf {\bibinfo {volume} {103}},\ \bibinfo {pages} {094007} (\bibinfo {year} {2021})},\ \Eprint {https://arxiv.org/abs/2104.01047} {arXiv:2104.01047 [hep-ph]} \BibitemShut {NoStop}%
\bibitem [{\citenamefont {Roca}\ and\ \citenamefont {Oset}(2010)}]{Roca:2010tf}%
  \BibitemOpen
  \bibfield  {author} {\bibinfo {author} {\bibfnamefont {L.}~\bibnamefont {Roca}}\ and\ \bibinfo {author} {\bibfnamefont {E.}~\bibnamefont {Oset}},\ }\bibinfo {title} {{Description of the $f_2(1270)$, $\rho_3(1690)$, $f_4(2050)$, $\rho_5(2350)$ and $f_6(2510)$ resonances as multi-$\rho(770)$ states}},\ \href {https://doi.org/10.1103/PhysRevD.82.054013} {\bibfield  {journal} {\bibinfo  {journal} {Phys. Rev. D}\ }\textbf {\bibinfo {volume} {82}},\ \bibinfo {pages} {054013} (\bibinfo {year} {2010})},\ \Eprint {https://arxiv.org/abs/1005.0283} {arXiv:1005.0283 [hep-ph]} \BibitemShut {NoStop}%
\bibitem [{\citenamefont {Boffi}\ {\it et~al.}(1991)\citenamefont {Boffi}, \citenamefont {Bracci},\ and\ \citenamefont {Christillin}}]{Boffi:1991nh}%
  \BibitemOpen
  \bibfield  {author} {\bibinfo {author} {\bibfnamefont {S.}~\bibnamefont {Boffi}}, \bibinfo {author} {\bibfnamefont {L.}~\bibnamefont {Bracci}},\ and\ \bibinfo {author} {\bibfnamefont {P.}~\bibnamefont {Christillin}},\ }\bibinfo {title} {{Coherent $(\gamma,\,\pi^0)$ on nuclei}},\ \href {https://doi.org/10.1007/BF02820558} {\bibfield  {journal} {\bibinfo  {journal} {Nuovo Cim. A}\ }\textbf {\bibinfo {volume} {104}},\ \bibinfo {pages} {843} (\bibinfo {year} {1991})}\BibitemShut {NoStop}%
\bibitem [{\citenamefont {Carrasco}\ {\it et~al.}(1993)\citenamefont {Carrasco}, \citenamefont {Nieves},\ and\ \citenamefont {Oset}}]{Carrasco:1991we}%
  \BibitemOpen
  \bibfield  {author} {\bibinfo {author} {\bibfnamefont {R.~C.}\ \bibnamefont {Carrasco}}, \bibinfo {author} {\bibfnamefont {J.}~\bibnamefont {Nieves}},\ and\ \bibinfo {author} {\bibfnamefont {E.}~\bibnamefont {Oset}},\ }\bibinfo {title} {{Coherent $(\gamma, \pi^0)$ photoproduction in a local approximation to the delta hole model}},\ \href {https://doi.org/10.1016/0375-9474(93)90005-I} {\bibfield  {journal} {\bibinfo  {journal} {Nucl. Phys. A}\ }\textbf {\bibinfo {volume} {565}},\ \bibinfo {pages} {797} (\bibinfo {year} {1993})}\BibitemShut {NoStop}%
\bibitem [{\citenamefont {Yamagata-Sekihara}\ {\it et~al.}(2011)\citenamefont {Yamagata-Sekihara}, \citenamefont {Nieves},\ and\ \citenamefont {Oset}}]{Yamagata-Sekihara:2010kpd}%
  \BibitemOpen
  \bibfield  {author} {\bibinfo {author} {\bibfnamefont {J.}~\bibnamefont {Yamagata-Sekihara}}, \bibinfo {author} {\bibfnamefont {J.}~\bibnamefont {Nieves}},\ and\ \bibinfo {author} {\bibfnamefont {E.}~\bibnamefont {Oset}},\ }\bibinfo {title} {{Couplings in coupled channels versus wave functions in the case of resonances: application to the two $\Lambda(1405)$ states}},\ \href {https://doi.org/10.1103/PhysRevD.83.014003} {\bibfield  {journal} {\bibinfo  {journal} {Phys. Rev. D}\ }\textbf {\bibinfo {volume} {83}},\ \bibinfo {pages} {014003} (\bibinfo {year} {2011})},\ \Eprint {https://arxiv.org/abs/1007.3923} {arXiv:1007.3923 [hep-ph]} \BibitemShut {NoStop}%
\bibitem [{\citenamefont {Martinez~Torres}\ {\it et~al.}(2011)\citenamefont {Martinez~Torres}, \citenamefont {Garzon}, \citenamefont {Oset},\ and\ \citenamefont {Dai}}]{MartinezTorres:2010ax}%
  \BibitemOpen
  \bibfield  {author} {\bibinfo {author} {\bibfnamefont {A.}~\bibnamefont {Martinez~Torres}}, \bibinfo {author} {\bibfnamefont {E.~J.}\ \bibnamefont {Garzon}}, \bibinfo {author} {\bibfnamefont {E.}~\bibnamefont {Oset}},\ and\ \bibinfo {author} {\bibfnamefont {L.~R.}\ \bibnamefont {Dai}},\ }\bibinfo {title} {{Limits to the Fixed Center Approximation to Faddeev equations: the case of the $\phi(2170)$}},\ \href {https://doi.org/10.1103/PhysRevD.83.116002} {\bibfield  {journal} {\bibinfo  {journal} {Phys. Rev. D}\ }\textbf {\bibinfo {volume} {83}},\ \bibinfo {pages} {116002} (\bibinfo {year} {2011})},\ \Eprint {https://arxiv.org/abs/1012.2708} {arXiv:1012.2708 [hep-ph]} \BibitemShut {NoStop}%
\bibitem [{\citenamefont {Toledo}\ {\it et~al.}(2021)\citenamefont {Toledo}, \citenamefont {Ikeno},\ and\ \citenamefont {Oset}}]{Toledo:2020zxj}%
  \BibitemOpen
  \bibfield  {author} {\bibinfo {author} {\bibfnamefont {G.}~\bibnamefont {Toledo}}, \bibinfo {author} {\bibfnamefont {N.}~\bibnamefont {Ikeno}},\ and\ \bibinfo {author} {\bibfnamefont {E.}~\bibnamefont {Oset}},\ }\bibinfo {title} {{Theoretical study of the $D^0 \to K^- \pi^+ \eta$ reaction}},\ \href {https://doi.org/10.1140/epjc/s10052-021-09058-z} {\bibfield  {journal} {\bibinfo  {journal} {Eur. Phys. J. C}\ }\textbf {\bibinfo {volume} {81}},\ \bibinfo {pages} {268} (\bibinfo {year} {2021})},\ \Eprint {https://arxiv.org/abs/2008.11312} {arXiv:2008.11312 [hep-ph]} \BibitemShut {NoStop}%
\bibitem [{\citenamefont {Jia}\ {\it et~al.}(2026{\natexlab{c}})\citenamefont {Jia}, \citenamefont {Su}, \citenamefont {Liang}, \citenamefont {Molina},\ and\ \citenamefont {Oset}}]{Jia:2025obs}%
  \BibitemOpen
  \bibfield  {author} {\bibinfo {author} {\bibfnamefont {W.-H.}\ \bibnamefont {Jia}}, \bibinfo {author} {\bibfnamefont {P.-S.}\ \bibnamefont {Su}}, \bibinfo {author} {\bibfnamefont {W.-H.}\ \bibnamefont {Liang}}, \bibinfo {author} {\bibfnamefont {R.}~\bibnamefont {Molina}},\ and\ \bibinfo {author} {\bibfnamefont {E.}~\bibnamefont {Oset}},\ }\bibinfo {title} {{Superexotic $K^{*+}D^{*+}K^{*+}$ bound state}},\ \href {https://doi.org/10.1016/j.physletb.2026.140320} {\bibfield  {journal} {\bibinfo  {journal} {Phys. Lett. B}\ }\textbf {\bibinfo {volume} {875}},\ \bibinfo {pages} {140320} (\bibinfo {year} {2026}{\natexlab{c}})},\ \Eprint {https://arxiv.org/abs/2512.04001} {arXiv:2512.04001 [hep-ph]} \BibitemShut {NoStop}%
\end{thebibliography}%
\end{document}